\documentclass[11pt,a4paper]{article}
\usepackage[margin=1in]{geometry}
\usepackage{amsmath,amssymb,amsthm}
\usepackage{graphicx}
\usepackage{booktabs}
\usepackage[round, authoryear]{natbib}
\usepackage{hyperref}
\usepackage{authblk}
\usepackage{caption}
\usepackage{float}
\usepackage{comment}

\title{Is Trend Still Your Friend?\\
\large A Microstructural Account of the Demise of Short-Term Trend-Following }

\author[1, 2]{Jutta G. Kurth}
\author[3]{Zoltan Eisler}
\author[4]{Adam Rej}
\author[4,1,5]{Jean-Philippe Bouchaud}
\affil[1]{Econophysics Lab, Institut Louis Bachelier, 28 Place de la Bourse, 75002 Paris, France}
\affil[2]{LadHyX UMR CNRS 7646, École polytechnique, 91128 Palaiseau Cedex, France}
\affil[3]{Imperial College London, Department of Mathematics}
\affil[4]{Capital Fund Management, 23 Rue de l'Université, 75007 Paris, France}
\affil[5]{Académie des Sciences, 23 Quai de Conti, 75006 Paris, France}

\date{\today}

\begin{document}
\maketitle

\begin{abstract}
\noindent
Systematic trend following has, on average, been profitable for at least
two centuries; yet since approximately 2009, short-term trends have
ceased to deliver reliable returns. Using a cross-section of roughly 100
liquid futures contracts spanning 1995--2025, together with an
industry-representative CTA proxy, we document the break and characterise
its dependence on signal speed and asset class. We evaluate four
candidate explanations -- capacity constraints, market electronification,
a regime change in CTA-versus-order-flow interactions, and a
microstructural mechanism -- and find that the first three fail on
grounds of timing, magnitude, or cross-sectional heterogeneity.

Our central empirical finding is that the cross-sectional variable
distinguishing degraded from surviving trends is the
\emph{volatility-normalised tick size}: post-2008 trend PnL has collapsed
on small-tick contracts across all signal horizons, while remaining
essentially intact on large-tick contracts. Neither asset class nor
liquidity replicates this dichotomy.

We interpret this result through the lens of a self-fulfilling feedback loop that, in our view, lies at the heart of the trend anomaly itself: \emph{Trend signals trigger directional trades, whose market impact reinforces the very price moves that generated the signal}. The profitability and the persistence of trend are therefore both sustained by the same impact channel. This loop, however, requires that trend followers can actually execute aggressively at reasonable cost. We argue that the post-crisis transition to HFT-dominated market making, whose liquidity-withdrawal behaviour in the face of predictable directional flow has sharply contrasting consequences for sparse (small-tick) and dense (large-tick) limit order books, has broken this loop on small-tick contracts. On large-tick contracts, where residual depth remains sufficient, the loop continues to operate and trend continues to deliver. A complementary price-impact analysis shows that passive execution offers no escape, since it forfeits the self-reinforcement channel while incurring adverse opportunity costs.

\end{abstract}

\section{Introduction}
\label{sec:intro}

Few empirical regularities in finance have proven as persistent and as
puzzling as the tendency of asset prices to trend. The observation that
recent winners outperform recent losers contradicts the most basic tenet
of the Efficient Market Hypothesis of \cite{fama1970efficient} -- obvious
information (past trends) is not incorporated into prices. Yet the
statistical evidence is overwhelming, spanning virtually every liquid
market and stretching back at least two centuries, as shown by
\cite{lemperiere2014two,hurst2017century}. \cite{moskowitz2012} document
``time-series momentum'' across 58 instruments and three and a half decades;
market-neutral momentum in the stock space is also well known
\citep{jegadeesh1993returns}; further work by \cite{baltas2013momentum},
\cite{levine2016trend}, and others has solidified trend following as
one of the most robust anomalies ever documented in finance.

The commercial exploitation of this anomaly through the Commodity
Trading Advisor (CTA) industry took off in the 1970s and 1980s, and by
the 1990s had converged on a fairly homogeneous methodology: take
positions proportional to the difference between a fast and a slow
moving average of prices, normalised by volatility, diversified across
as many liquid futures contracts as possible.

\begin{figure}[h]
\centering
\includegraphics[width=0.85\textwidth]{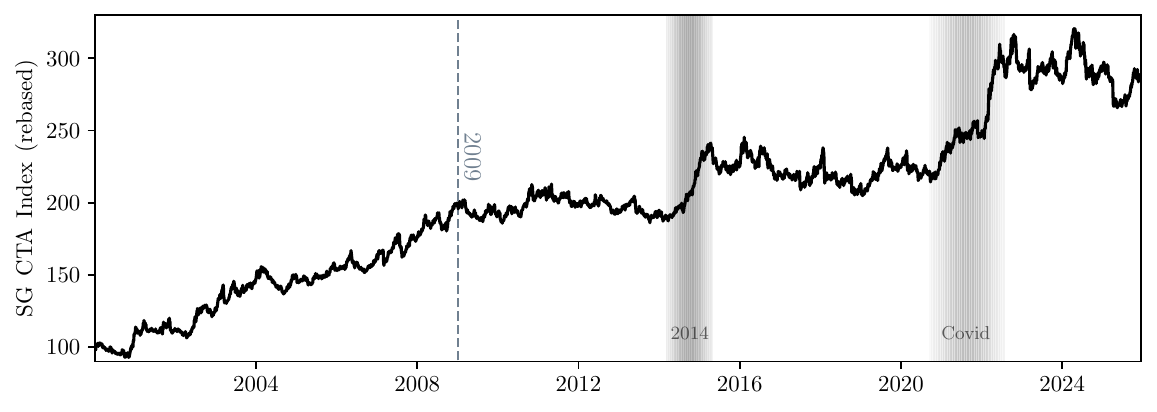}
\caption{(Rebased) SG CTA Index, 2000--2025. Data courtesy of \cite{SG_STF_Index}.}
\label{fig:sg_index}
\end{figure}

And yet, since approximately 2008/9, this anomaly appears to have been
struggling, at least when looking at the performance of the SG CTA
Index (the industry-standard benchmark for large CTAs) from 2000 to
2025; see Fig.~\ref{fig:sg_index}. The contrast between the pre- and
post-2008/9 regimes is stark: after nearly a decade of steady gains,
performance has been effectively flat or negative for the past fifteen
years, only saved by two macro-driven punctuations in 2014 and during
Covid. A clean methodological proxy that we construct below paints
the same picture with both sharper resolution and more nuance.

Several stylised facts can be established and constrain any candidate
explanation: (i) the break is \emph{abrupt}, with the character of a
structural regime shift rather than gradual alpha decay; (ii) it is
\emph{speed-dependent}, with fast signals (horizons of days to weeks)
most affected whereas slow ones are approximately intact; (iii) it is
\emph{asset-class heterogeneous}, sparing yields and most commodities
(in our universe) while strongly depressing equity indices and
currencies; and (iv) there has been \emph{no recovery}, despite material
improvements in liquidity and declines in CTA participation since 2018.

\paragraph{The self-fulfilling impact loop.}
Before turning to the post-2008 break, it is useful to articulate what
trend, viewed as a market mechanism rather than as a statistical
anomaly, actually consists of. We adopt throughout this paper the view
-- developed precisely in the ETF context by \cite{van2024ponzi}, and
foreshadowed by \cite{deLong1990positivefeedback} -- that trend 
is chiefly (but most probably not only, see below) a self-fulfilling feedback loop:
\begin{quote}
\emph{trend signals trigger directional trades, whose market impact
reinforces the very price moves that produced the signal, which in turn
sustains the signal for the next trader to act on.}
\end{quote}
The profitability of trend and its very existence are thus two faces of
the same coin: both depend on aggressive directional flow being
absorbable by the market at reasonable cost, so that impact translates
trade into price and price into renewed signal. Anything that breaks
this loop -- by raising execution costs to the point of disengagement,
or by altering the relationship between aggressive flow and price --
should attack profitability and signal strength simultaneously. This is
the lens through which we read the post-2008 evidence.

\paragraph{Other mechanisms exist; the loop is not exclusive.}
We do not claim that the self-fulfilling impact loop is the sole
origin of trend. The behavioural-finance literature offers
well-established alternatives -- notably gradual incorporation of
public news under underreaction
\citep{hong1999unified, daniel1998investor, bouchaud2019sticky}, the slow
diffusion of information across heterogeneously informed investors
\citep{hong2000bad}, and staggered accumulation by
participants correctly anticipating a future event -- each of which
can produce price autocorrelation without invoking impact-mediated
feedback. These mechanisms are not mutually exclusive with the loop;
they plausibly coexist, with information-diffusion channels seeding
trends and the impact loop amplifying them.  In fact, we have
direct empirical evidence in our data that some traders did
anticipate large market moves and create trends accordingly, and
that this channel has itself decayed around approximately the same
period -- see Appendix~\ref{app:decompositions}, Fig.~\ref{fig:PnL_STLT_byReturnMagnitude}. The empirical
observations we report below are therefore most naturally read as
evidence that \emph{both} the impact-loop and the
anticipatory-trading components have been weakened in the post-2008
period, with the loop component decaying most visibly on small-tick
contracts. We adopt the loop as our organising lens because it
makes the cleanest predictions about the cross-section we study,
not because we believe it is the only mechanism at work.

\paragraph{Plan of attack.}
This paper assesses four candidate explanations: (H1) capacity
constraints; (H2) the electronification of futures markets; (H3) a
structural shift in the interaction between CTA trades and aggregate
order flow; and (H4) a microstructural mechanism whereby liquidity
offered to trend followers has dried up. We reject H1--H3 on the grounds
of timing, magnitude, or cross-sectional heterogeneity. Our central
contribution concerns H4: the relevant cross-sectional variable
separating degraded from surviving trend strategies is the
\emph{volatility-normalised tick size}. Partitioning the futures
universe by this variable produces a clean dichotomy that no
decomposition by asset class or liquidity replicates.

We propose a mechanism rooted in the post-crisis transition to
HFT-dominated market making, whose liquidity-withdrawal behaviour in
front of predictable directional flow has markedly different
consequences for sparse (small-tick) and dense (large-tick) order
books. Combined with the self-fulfilling loop above, this mechanism
predicts -- and the data confirm -- that the small-tick collapse is
not merely a story of higher transaction costs eating into PnL, but
of the impact-mediated reinforcement of trends having been severed
on those contracts. A complementary price-impact analysis shows that
passive execution offers no escape: it forfeits the self-reinforcement
channel while incurring adverse book-impact.

The paper is organised as follows. Sec.~\ref{sec:portfolio} defines
the trend signals, portfolios, and CTA proxy. Sec.~\ref{sec:decay}
documents the empirical decay across speed and asset class.
Sec.~\ref{sec:H1H2H3} assesses H1--H3. Sec.~\ref{sec:ticksize}
establishes tick size as the relevant cross-sectional variable.
Sec.~\ref{sec:mechanism} develops the microstructural mechanism,
including the role of price impact and the self-fulfilling loop.
Sec.~\ref{sec:conclusion}
concludes. Further supplements and details are provided in various
appendices.

\section{Trend Portfolios and a CTA Industry Proxy}
\label{sec:portfolio}

We work with a cross-section of approximately 100 of the most liquid
futures contracts, sampled at daily settlement, spanning four sectors:
commodities (CMD), equity indices (IDX), currencies (FXR), and
government bonds/yields (YLD). The data catalogue is detailed in
Appendix~\ref{app:data}.

\paragraph{Signal definition.}
For each contract $i$ and trading day $t$, the unnormalised trend
signal on time scale $\tau$ is the difference between a fast and a
four-times slower exponentially weighted moving average (EWMA) of
close prices,
$\tilde{s}_i^{\tau}(t) = \langle p_i(t-1)\rangle_\tau
- \langle p_i(t-1)\rangle_{4\tau}$.
Normalising by a slow estimate of its own dispersion yields the
dimensionless signal
\begin{equation}
s_i^\tau(t) = \frac{\tilde{s}_i^\tau(t)}{\langle\langle
\tilde{s}_i^\tau(t)\rangle\rangle_{16\tau}},
\label{eq:signal}
\end{equation}
which is comparable across contracts. Here $\langle \cdot \rangle_\tau$
refers to the backward-looking EWMA with characteristic decay time
$\tau$ (in days) and $\langle\langle \cdot \rangle\rangle_\tau$ refers
to an exponentially weighted moving standard deviation (EWMSTD). We
refer to a signal with parameter $\tau$ as EWM-$\tau$-$4\tau$. As is
industry practice, signals are clipped at $\pm 2$ standard deviations:
$s_i^\tau(t) \mapsto \mathrm{clip}(s_i^\tau(t), -2, 2)$.

\paragraph{Portfolio construction.} Two volatility-scaled portfolios
constructed from all contracts are:
\begin{itemize}
    \item[]\textbf{\hspace{-0.6cm}Type 1 (equal-risk):} the position
    is volatility-normalised,
    \begin{equation}
        \pi_i^\tau (t) = \frac{s_i^\tau (t)}{\sigma_i (t)}.
        \label{eq:portflio_wo_liq_weighting}
    \end{equation}
    \item[] \textbf{\hspace{-0.6cm}Type 2 (liquidity-weighted):} the
    position is normalised by its own volatility and by its liquidity
    expressed as a ratio of the aggregate liquidity in the portfolio,
    \begin{equation}
        \pi_i^\tau (t) = \frac{s_i^\tau (t)\, l_i (t)}
        {\sigma_i (t)\, L (t)}.
        \label{eq:portfolio_liqw}
    \end{equation}
\end{itemize}
The volatility is defined as
$\sigma_i(t) = \langle |p_i(t-1) - p_i(t-2)| \rangle_{16\tau}$;
$l_i(t) = \langle \text{V}_i^{\text{w/o imp}}(t-1)\, \sigma_i(t)
\rangle_{250}$ is a contract's liquidity, and
$L (t) = \sum_i l_i (t)$ is the aggregate portfolio liquidity, where
$\text{V}_i^{\text{w/o imp}}$ is the traded volume in contract $i$
without implied trades. On each trading day the position is updated
with the trade
$\Delta\pi_i^{\tau}(t) = \pi_i^{\tau}(t) - \pi_i^{\tau}(t-1)$
per product. The volatility normalisation ensures that each contract
contributes a similar dollar-risk to the overall portfolio, while the
liquidity-weighting becomes essential when scaling up to (CTA)
industry size, since illiquid products cannot be scaled up in the
same way as very liquid ones owing to differing price-impact costs.

\paragraph{CTA proxy.} To match the risk of the aggregate CTA
industry, the liquidity-weighted portfolio is rescaled according to
industry-wide AUM $S^{\text{CTA}}(t)$ from
\citet{BarcleyHedge_CTA_AUM} at an annualised risk of
$r_a \approx 12\%$ (inferred from the SG CTA index):
\begin{equation}
\pi_i^{\text{CTA},\tau}(t) = \pi_i^\tau(t)\,
\frac{S^{\text{CTA}}(t)}{\$1}\,
\frac{r_a}{\sqrt{252}}.
\label{eq:scaleup}
\end{equation}
Computing the resulting daily participation rate
$\text{PR}_i = |\Delta\pi_i^{\text{CTA}}|/\text{V}_i^{\text{w/o imp}}$
across contracts gives a cross-sectional median of $\sim$0.9\%,
broadly consistent with independent estimates by
\citet{quantica2022trend_footprint} and
\citet{MorganStanley2025CTA}: the CTA industry represents less than
$1\%$ of the exchanged volume on futures. This agreement provides
external validation of the proxy in Eq.~\eqref{eq:scaleup}.

\section{Empirical Decay of Trend PnL}
\label{sec:decay}

Equipped with the trend portfolios and the CTA proxy of
Sec.~\ref{sec:portfolio}, we now document the post-2008
performance break empirically. Three features emerge that any
viable explanation must accommodate: an abrupt aggregate decay
from 2008/9 onward, a strong dependence on signal speed (with fast
trends most affected), and a marked heterogeneity across asset
classes. Each is the subject of a dedicated subsection below. We
present these patterns descriptively here; their mechanistic
interpretation, and in particular the role of the self-fulfilling
impact loop, is deferred to Sec.~\ref{sec:mechanism}.

\subsection{Aggregate cumulative PnL} 

\begin{figure}[h]
\centering
\includegraphics[width=0.9\textwidth]{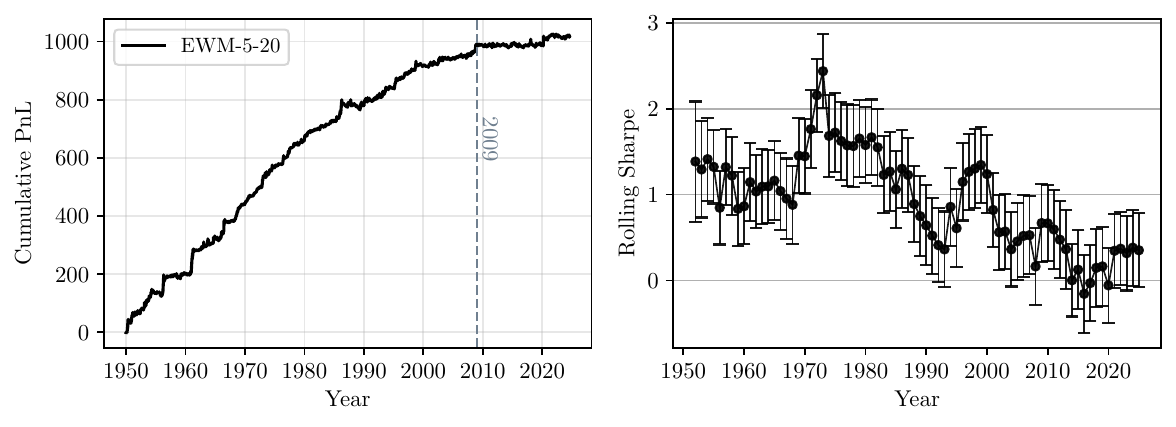}
\caption{Left: cumulative PnL of EWM-5-20, 1950--2025.
Right: corresponding 5-year rolling annualised Sharpe with bootstrap
standard errors.}
\label{fig:pnl_5_20}
\end{figure}

Fig.~\ref{fig:pnl_5_20} (left) shows the
cumulative PnL of a fast trend portfolio (EWM-5-20) from 1950 to
2025. Two features stand out: the cumulative PnL is essentially flat
from 2009 onward, although a weakening may already be detected from
approximately 2000. The five-year rolling Sharpe ratio (right) collapses
from a historical range of 1--2.5 to a level statistically
indistinguishable from zero post-2010. This was also reported in
\cite{lemperiere2014two}, and more recently by
\cite{schmidhuber2021trend_reversion}.

\subsection{Speed dependence}
\begin{figure}[htbp]
\centering
\includegraphics[width=0.9\textwidth]{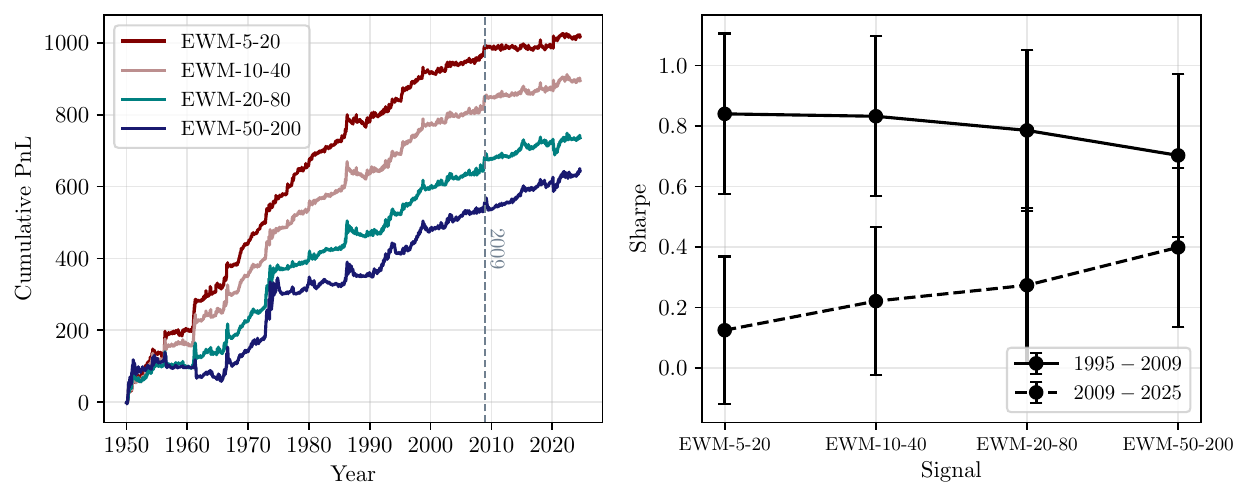}
\caption{Left: cumulative PnL of EWM-$\tau$-$4\tau$ portfolios for a
range of fast scales $\tau$. Right: corresponding pre- and post-break
Sharpe by signal. See also Table~\ref{tab:sharpe_speed}.}
\label{fig:PnL_Sharpe_speed_class}
\end{figure}

Repeating the analysis for $\tau \in \{5,10,20,50\}$ reveals a
striking asymmetry (Table~\ref{tab:sharpe_speed} and
Fig.~\ref{fig:PnL_Sharpe_speed_class}): pre-2009 Sharpes are
monotonically decreasing in $\tau$ (from 0.84 at $\tau{=}5$ to 0.70
at $\tau{=}50$), whereas post-2008 the ordering reverses, with the
fastest signal collapsing to 0.12 and the slowest still delivering
0.40.

\begin{table}[h]
\centering
\begin{tabular}{ccc}
\toprule
$\tau$ & Sharpe 1995--2009 & Sharpe 2009--2025 \\
\midrule
5  & 0.84 $\pm$ 0.27 & 0.12 $\pm$ 0.24\\
10 & 0.83 $\pm$ 0.27 & 0.22 $\pm$ 0.24 \\
20 & 0.79 $\pm$ 0.27 & 0.27 $\pm$ 0.26 \\
50 & 0.70 $\pm$ 0.27 & 0.40 $\pm$ 0.26 \\
\bottomrule
\end{tabular}
\caption{Sharpe ratios of EWM-$\tau$-$4\tau$ trend signals over two
sub-periods straddling the performance break. See also
Fig.~\ref{fig:PnL_Sharpe_speed_class}.}
\label{tab:sharpe_speed}
\end{table}

\subsection{Asset-class heterogeneity}
\begin{figure}[h]
\centering
\includegraphics[width=0.55\textwidth]{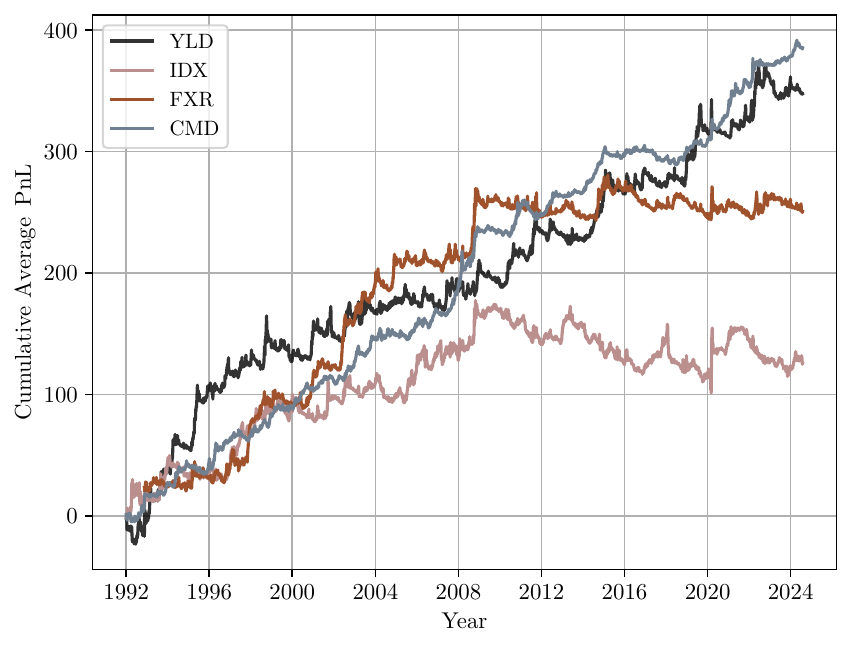}
\caption{Cumulative PnL of a fast strategy (EWM-5-20), averaged
within each asset class.}
\label{fig:PnL_byAC}
\end{figure}

Fig.~\ref{fig:PnL_byAC} disaggregates the cumulative PnL by sector.
Trend has effectively vanished for IDX and FXR, while YLD and CMD
show no appreciable degradation. As we will see in Sec.~\ref{sec:ticksize}, this asset-class heterogeneity is most cleanly understood as a downstream consequence of a deeper microstructural variable.

\section{Assessing Capacity, Electronification, and Order Flow}
\label{sec:H1H2H3}

Having documented the post-2008 break and its qualitative
signatures, we now confront three of the most natural candidate
explanations: capacity constraints (H1), the electronification of
futures markets (H2), and a structural shift in the interaction
between CTA trades and aggregate order flow (H3). For each, we ask
whether the timing, the magnitude, and the cross-sectional pattern
of the proposed channel match those of the observed PnL decay. We
find that none does, although H3 will turn out to be a useful
diagnostic when revisited at the right level of cross-section in
Sec.~\ref{sec:ticksize}.

\subsection{H1: Capacity constraints}

The naive crowding argument -- more capital arbitrages the signal
away -- is not viable for trend, since rational speculators in the
presence of positive feedback should front-run rather than weaken
trends \citep{deLong1990positivefeedback}. The relevant version is
therefore the execution-cost story articulated by
\citet{quantica2025trend_capacity}\footnote{Strictly speaking, \citet{quantica2025trend_capacity} do not argue that trend has hit capacity industry-wide or that capacity constraints explain the post-2008 break. Their analysis is cross-sectional: at large AUM, risk allocation must concentrate in the most liquid commodities, eroding the diversification benefit that less liquid markets would otherwise contribute (they estimate ~17\% Sharpe drag at \$1B commodity capacity). They explicitly find no relationship between per-instrument Sharpe and liquidity after costs. We invoke their work here as the most carefully quantified version of the capacity hypothesis available, and as a useful motivator for our subsequent microstructural analysis, while acknowledging that capacity-driven erosion of diversification benefits in less liquid contracts may well be a genuine secondary effect operating alongside the mechanism we identify.} (see also
\cite{volpati_eisler2020zooming}): under the square-root impact law,
total slippage scales as $Q^{3/2}$ (see, e.g.,
\cite{toth2011impact, bouchaud2018trades}), and beyond some position
size $Q$ net Sharpe collapses. We find this hypothesis problematic
on four grounds.

\begin{figure}[htbp]
\centering
\includegraphics[width=0.42\textwidth]{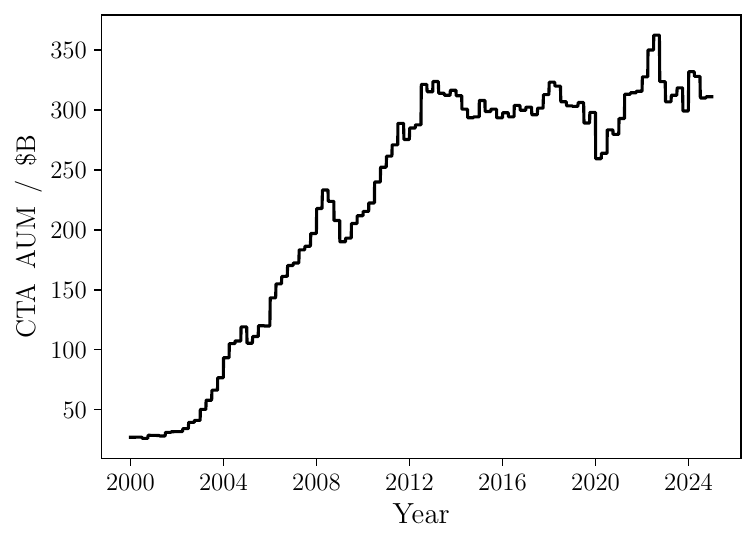}
\includegraphics[width=0.56\textwidth]{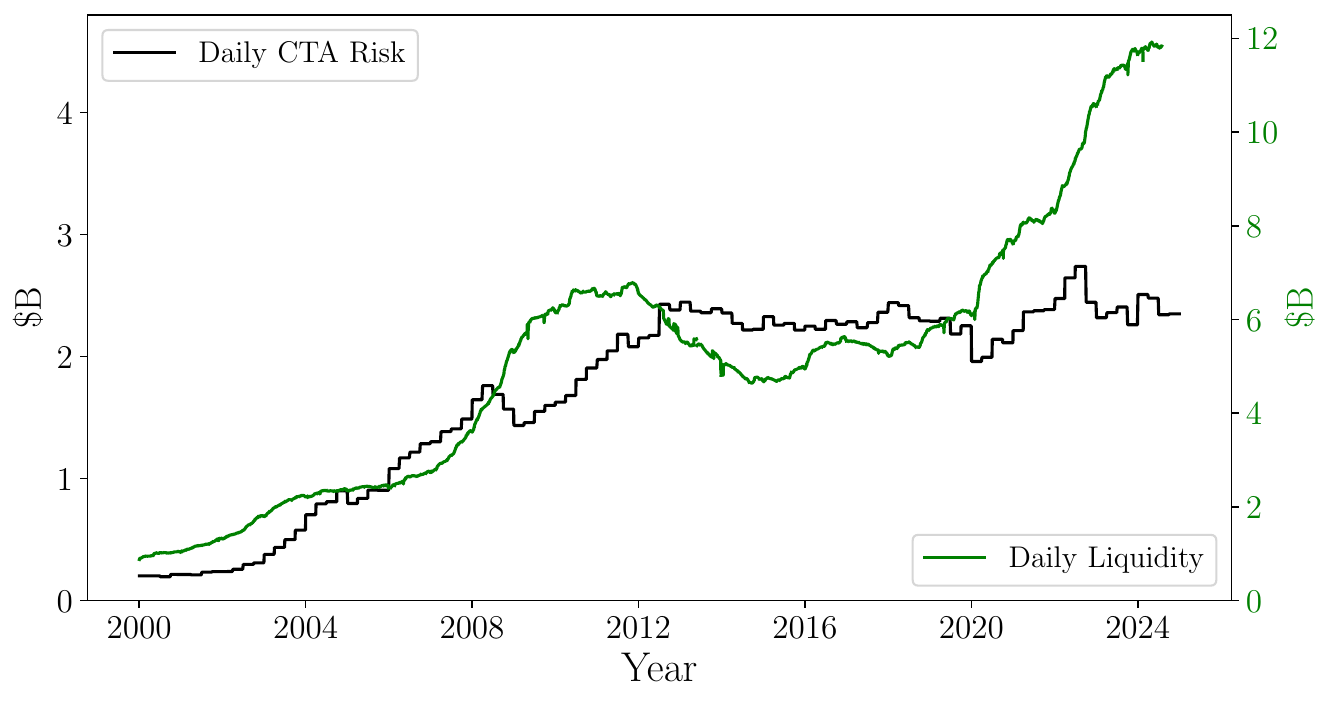}
\caption{Left: estimated CTA industry AUM, 2000--2025
\citep{BarcleyHedge_CTA_AUM}. Right: daily dollar risk of the CTA
industry at $r_a=12\%$ (black, left axis) and aggregate daily
liquidity (green, right axis).}
\label{fig:aum_liq}
\end{figure}

\emph{First, reverse causality.} CTA AUM (Fig.~\ref{fig:aum_liq})
grew through the 2000s, plateaued around 2012, and peaked in 2022 --
several years \emph{after} trend PnL went flat. The temporal ordering
is the opposite of what the capacity story predicts.

\emph{Second, no post-2018 recovery.} While CTA AUM stagnated after
2012, futures liquidity rose sharply from 2018 onward
(see Fig.~\ref{fig:aum_liq}, right), mechanically reducing
participation rates. Despite this, trend PnL did not recover.
Crowding-driven decay typically exhibits a recovery signature once
excess positions unwind \citep{mitchell2012arbitrage}; trend shows
none.

\emph{Third, the implied drag is too small.} A standard square-root
impact calculation (Appendix~\ref{app:capacity}) at a 1\%
participation rate and $\sim$9\% daily turnover yields an annualised
Sharpe drag of roughly 0.1. This is non-trivial but still too small
to explain the observed collapse from $\sim$0.7 to $\sim$0, even when
allowing for a factor-three error margin due to model uncertainty.

\noindent
\begin{minipage}{0.40\linewidth}\vspace{0.1cm}
\quad\;\emph{Fourth (and most telling), signal degradation under zero-lag execution.}
Recomputing the PnL with positions executed at the same day's close
(Fig.~\ref{fig:lag0}) -- mechanically eliminating one full day of
impact -- produces equally flat post-2008 returns, even without
costs. If capacity were the cause, removing execution costs should
restore profitability, which it does not. This means the signal
itself has degraded, rather than the strategy simply having become
too costly to trade. As we will see, this fourth observation is
particularly telling: it is the first hint that what has decayed
is not merely the harvest but the underlying signal -- exactly what
the self-fulfilling loop framework predicts when the loop is broken.
\end{minipage}%
\hfill
\begin{minipage}{0.58\linewidth}
     \centering
    \includegraphics[width=\linewidth]{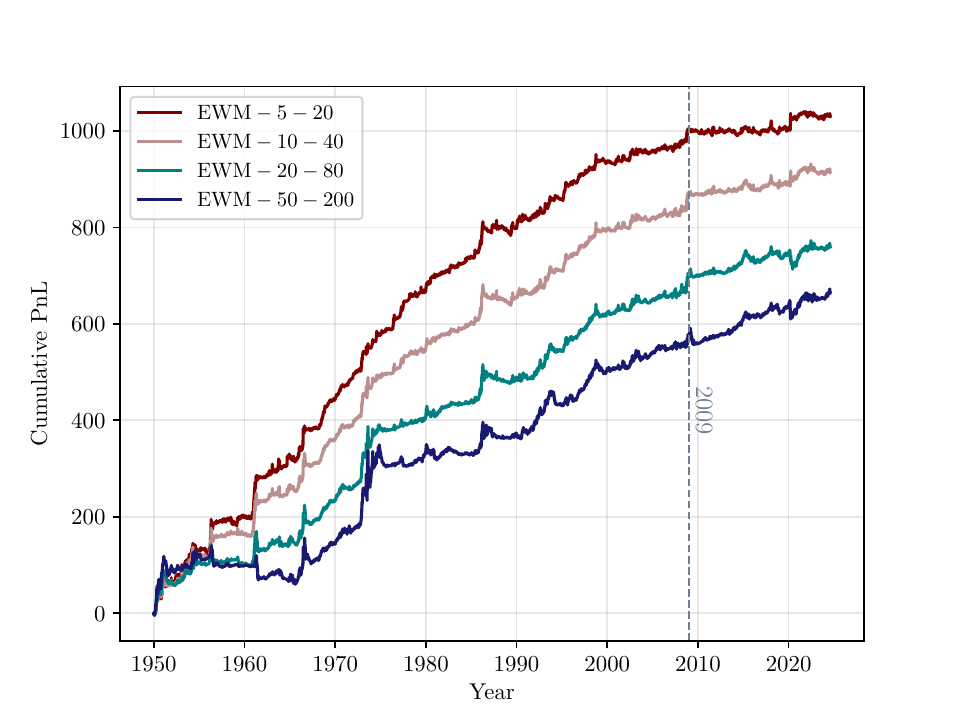}
    \captionof{figure}{Cumulative PnL with zero-lag execution: positions identified on day $t$ are executed at the day-$t$ close. Fast-signal performance (red, rosy) still remains flat post-2008.}
\label{fig:lag0}
\end{minipage}

\subsection{H2: Electronification}

\begin{minipage}{0.40\linewidth}
The transition to electronic order books on major futures exchanges
is sometimes invoked as a cause. Three observations from
Fig.~\ref{fig:electr_trading_fraction}, which shows the fraction of
electronically executed futures volume per sector on CME (Globex),
cast doubt on that story: electronification was \emph{gradual},
while the PnL break is abrupt; its sectoral \emph{timing} does not
align with the PnL break (equities electronified more than five
years before the break, FX about contemporaneously); and its
sectoral \emph{ordering} does not match the PnL response
(interest-rate futures electronified early but did not degrade).
Although market electronification is a necessary precondition for
the mechanism we propose below -- it concentrates flow into a single
visible book -- it is not by itself the cause.
\end{minipage}%
\hfill
\begin{minipage}{0.58\linewidth}
     \centering
    \includegraphics[width=0.9\textwidth]{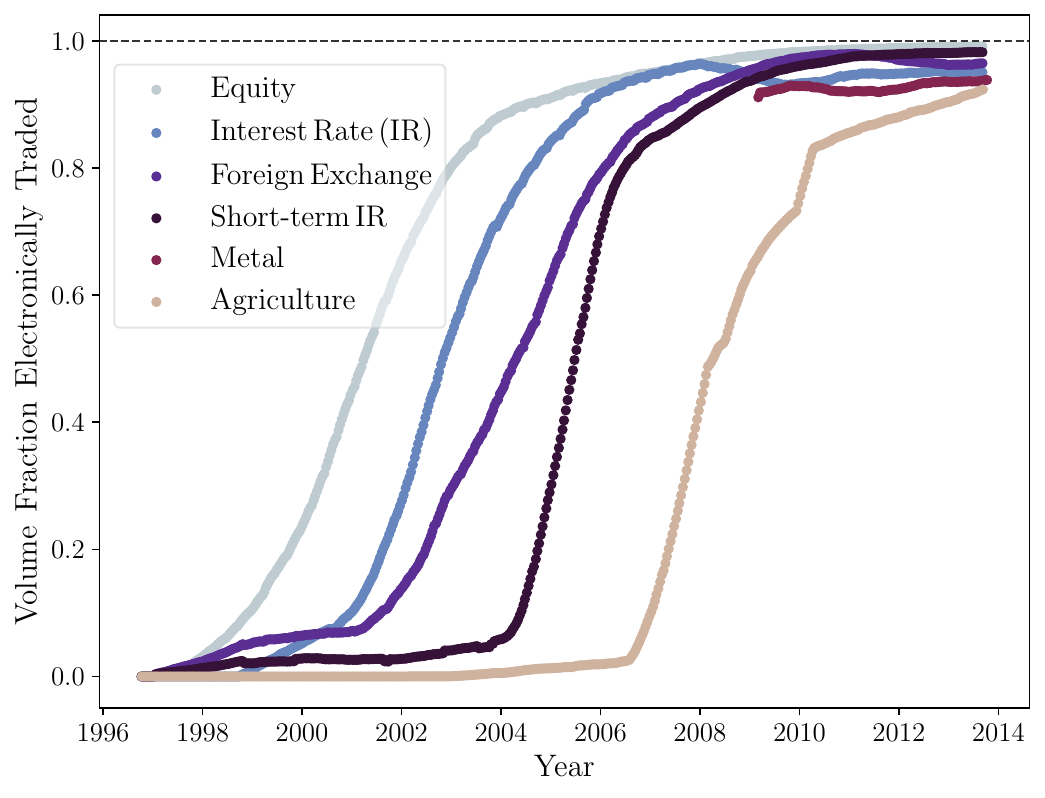}
    \captionof{figure}{Progression of the futures volume fraction by asset class that is traded electronically on Globex versus the total traded CME futures volume, which includes pit trading through open outcry. Data courtesy of Robert Almgren, Quantitative Brokers.}
    \label{fig:electr_trading_fraction}
\end{minipage}

\subsection{H3: Order-flow interactions}
\label{sec:H3_orderflow}

Following \citet{volpati_eisler2020zooming}, we define daily book
and trade imbalances on volume-weighted 5-minute bars:
\begin{equation}
I^{\text{book}}_{i,t} = \tfrac{1}{V_{i,t}}\!\sum_{s\in t}
\tfrac{V^{\text{bid}}_{i,s}-V^{\text{ask}}_{i,s}}
{V^{\text{bid}}_{i,s}+V^{\text{ask}}_{i,s}}\,V_{i,s},\quad
I^{\text{trade}}_{i,t} = \tfrac{1}{V_{i,t}}\!\sum_{s\in t}
\tfrac{V^{\text{BIV}}_{i,s}-V^{\text{SIV}}_{i,s}}
{V^{\text{BIV}}_{i,s}+V^{\text{SIV}}_{i,s}}\,V_{i,s},
\label{eq:imbalances}
\end{equation}
where $V_{i,s}$ is the volume traded in product $i$ in each of the
5-minute time windows in a trading day, and $V_{i,t}$ is the total
traded volume of contract $i$ in day $t$. $V^{\text{bid/ask}}$ is a
snapshot of the volume at the best bid/ask at the end of each bar
window, and $V^{\text{BIV/SIV}}$ is the cumulative
buyer/seller-initiated volume per interval. The volume-weighting
ensures that the daily statistics are dominated by liquid trading
hours and not diluted by overnight futures trading.

The idea is to study the correlation between these flow indicators
$I$ and the expected change of position of trend followers,
$\Delta \pi^{\text{CTA}}$, in order to detect possible changes in
the way trend followers trade and, symmetrically, in the liquidity
offered to them.

\begin{figure}[htbp]
    \centering
    \includegraphics[width=0.49\linewidth]{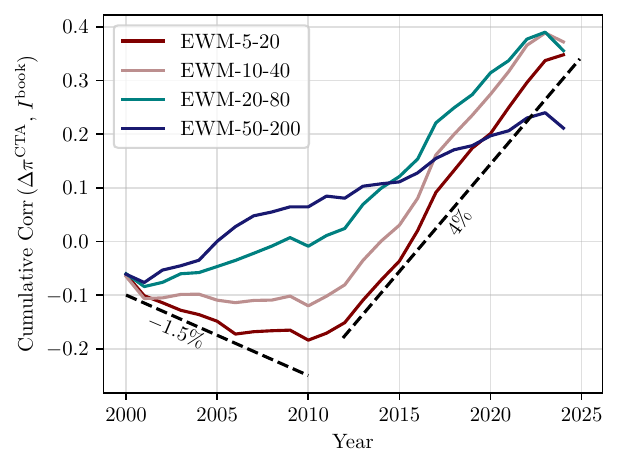}
    \includegraphics[width=0.49\linewidth]{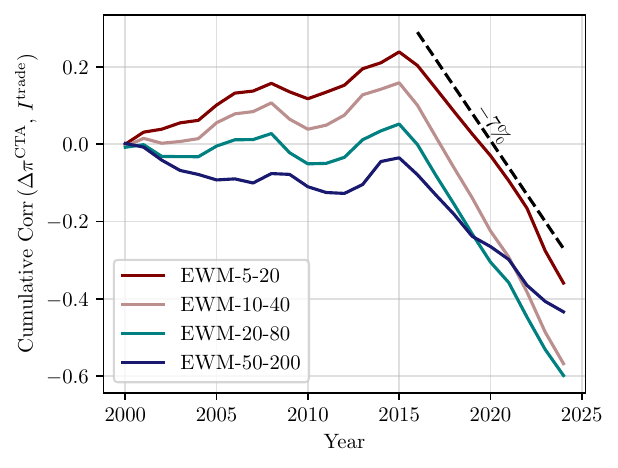}
    \caption{Annual cumulated correlations of trades
    $\Delta \pi^{\textup{CTA}, \tau}$ with book imbalances (left)
    and with trade imbalances (right), averaged over the entire
    portfolios built from different signals EWM-$\tau$-$4\tau$.
    Dashed lines are for guidance.}
    \label{fig:Corr_trades_Imbs_all}
\end{figure}

Aggregated across all futures markets, the correlation between
$\Delta \pi^{\text{CTA}}$ and book imbalance $I^{\text{book}}$ has
abruptly changed sign around 2010, especially for short-term trends
(5--20 days), while showing essentially no change for long-term
trends (Fig.~\ref{fig:Corr_trades_Imbs_all}, left). This is striking
because it precisely parallels the change in PnL of short trends
reported above, and motivates our initial working hypothesis. Indeed,
$\text{Corr}(\Delta\pi^{\text{CTA}}, I^{\text{book}})$ was negative
before 2010 and positive after. In other words, the order book was
thicker on the opposite side of the trend trade before 2010, meaning
that liquidity was provided to trend followers during that period.
Since then, liquidity has been more abundant for trades opposing
the trend, which may indicate that either liquidity providers shy
away from trading with trend followers, or that trend followers
themselves use more limit orders to execute -- or both.

The correlation with trade imbalance,
$\text{Corr}(\Delta\pi^{\text{CTA}}, I^{\text{trade}})$, on the
other hand, does not show any particular feature around 2010, but
does change sign abruptly in 2015, suggesting that ``anti-trend''
(i.e. mean-reversion) trades have become dominant in the recent
period, striking another blow to trend-following performance.

\begin{figure}[htb]
    \centering
    \includegraphics[width=0.86\linewidth]{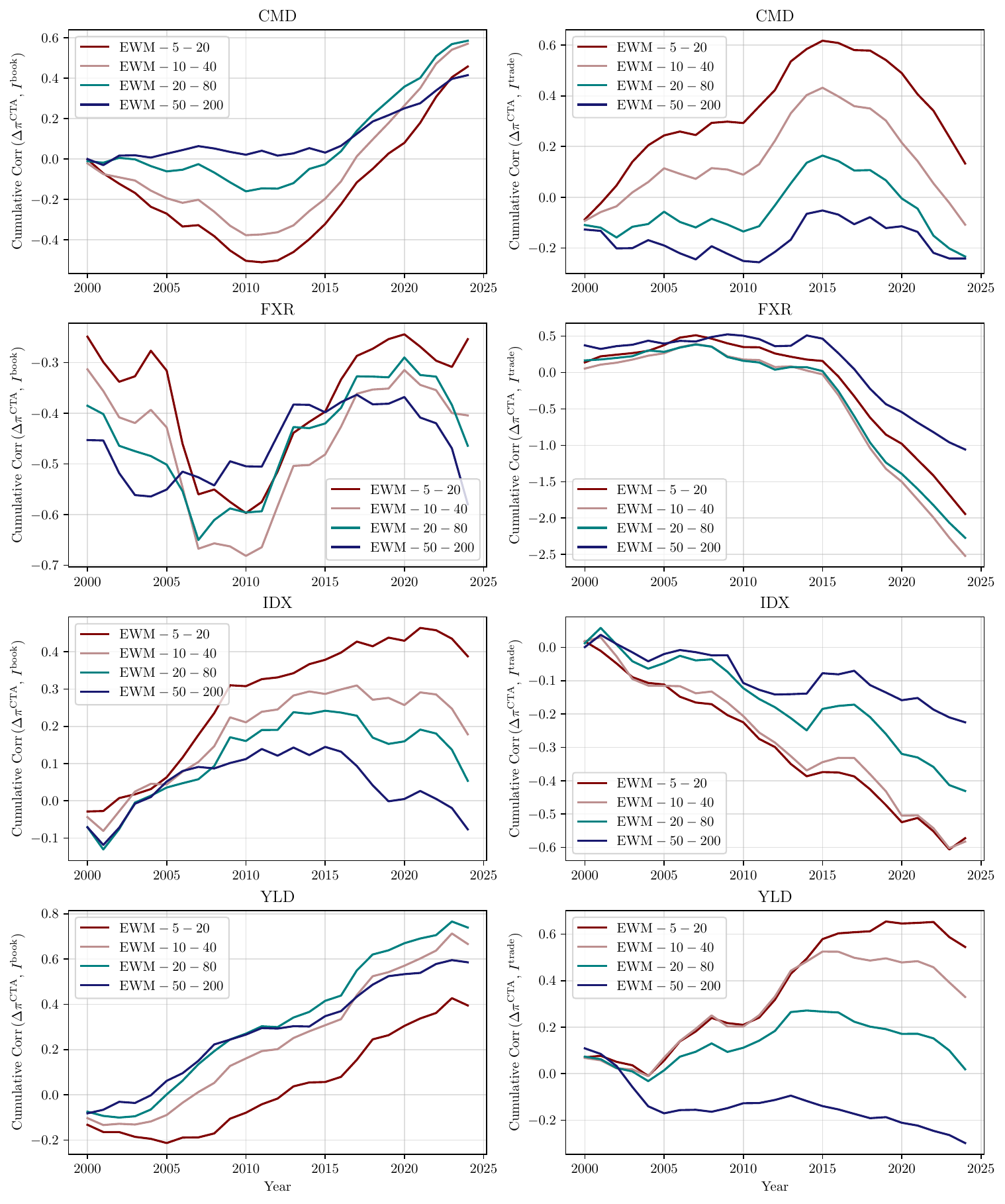}
    \caption{Same as Fig.~\ref{fig:Corr_trades_Imbs_all} but averaged
    per asset class.}
    \label{fig:corr_perAC}
\end{figure}

Although suggestive, this hypothesis fails the cross-sectional
test. Disaggregating by asset class, as shown in
Fig.~\ref{fig:corr_perAC}, reveals decisive mismatches between
correlation changes and PnL outcomes (see Fig.~\ref{fig:PnL_byAC}):
\begin{enumerate}
\item Commodities undergo a sharp book-imbalance regime change but
exhibit \emph{no} PnL degradation whatsoever; currencies show a
similar change in correlation structure and the post-2010 PnL is
flat.
\item Yields show essentially no correlation change and equally show
no PnL degradation.
\item Indices show a reverse pattern but a similar trend
degradation.
\item The post-2010 sign is common across YLD, CMD, and FXR but
their PnL outcomes differ entirely.
\end{enumerate}
No monotonic mapping exists between the correlation changes and the
PnL outcomes at the asset-class level. An intra-day volume-clock decomposition
(Appendix~\ref{app:volumeclock}) reveals that the first $20\%$ of daily volume behaves systematically differently from the rest -- consistent with morning-concentrated CTA execution -- but does not resolve the cross-sectional inconsistency.

H3 is therefore insufficient at
the asset-class level of decomposition. The diagnostic itself,
however, will re-emerge usefully at a different level of
cross-section under H4.

\section{Tick Size as a Discriminant Factor}
\label{sec:ticksize}

We now establish the central empirical result of the paper: the
relevant cross-sectional variable separating degraded from surviving
trend strategies is the \emph{volatility-normalised tick size}.

\subsection{Tiering procedure}
For each product $i$ and month $m$, define the average
tick-to-volatility ratio
\begin{equation}
\bar\rho_{i,m} = \frac{1}{D_m}\sum_{t=1}^{D_m}\frac{\Psi_i(t)}{\sigma_i(t)},
\label{eq:tick_vol_ratio}
\end{equation}
where $\Psi_i$ is the tick size of the contract, $\sigma_i$ is the
daily volatility estimator of Sec.~\ref{sec:portfolio}, and $D_m$
is the number of trading days in month $m$. Each month, products
are ranked in ascending order by $\bar\rho_{i,m}$,
\begin{equation}
    R_{i,m}
    :=
    \operatorname*{rank}_{j \in \mathcal{P}_{m}}
    \!\left( \bar{\rho}_{j,m} \right)\!\bigl[i\bigr]
    \;\in\; \{1, 2, \dots, N_{m}\},
    \label{eq:tick_rank}
\end{equation}
where $\mathcal{P}_{m}$ denotes the set of
$N_{m} \equiv |\mathcal{P}_{m}|$ products, and partitioned into two
equal-sized tiers:
\begin{equation}
    \mathrm{Tier}_{i,m}
    :=
    \begin{cases}
        \text{small tick (ST),}
            & \text{if } R_{i,m} \leq \lfloor N_{m} / 2 \rfloor, \\[4pt]
        \text{large tick (LT),}
            & \text{otherwise}.
    \end{cases}
    \label{eq:tick_tier}
\end{equation}
A product is categorised as small (large) tick if its
tick-to-volatility ratio is in the lowest (highest) 50\%. The
partition is used causally and recomputed monthly, so that a
contract whose volatility regime or tick size changes can migrate
between tiers. Effectively, however, products rarely migrate between tiers in our sample. Note that ST and LT refer to tick size, not to be confused with ``short trends'' or ``long trends''.

\subsection{Results}
\begin{figure}[htbp]
\centering
\includegraphics[width=0.95\textwidth]{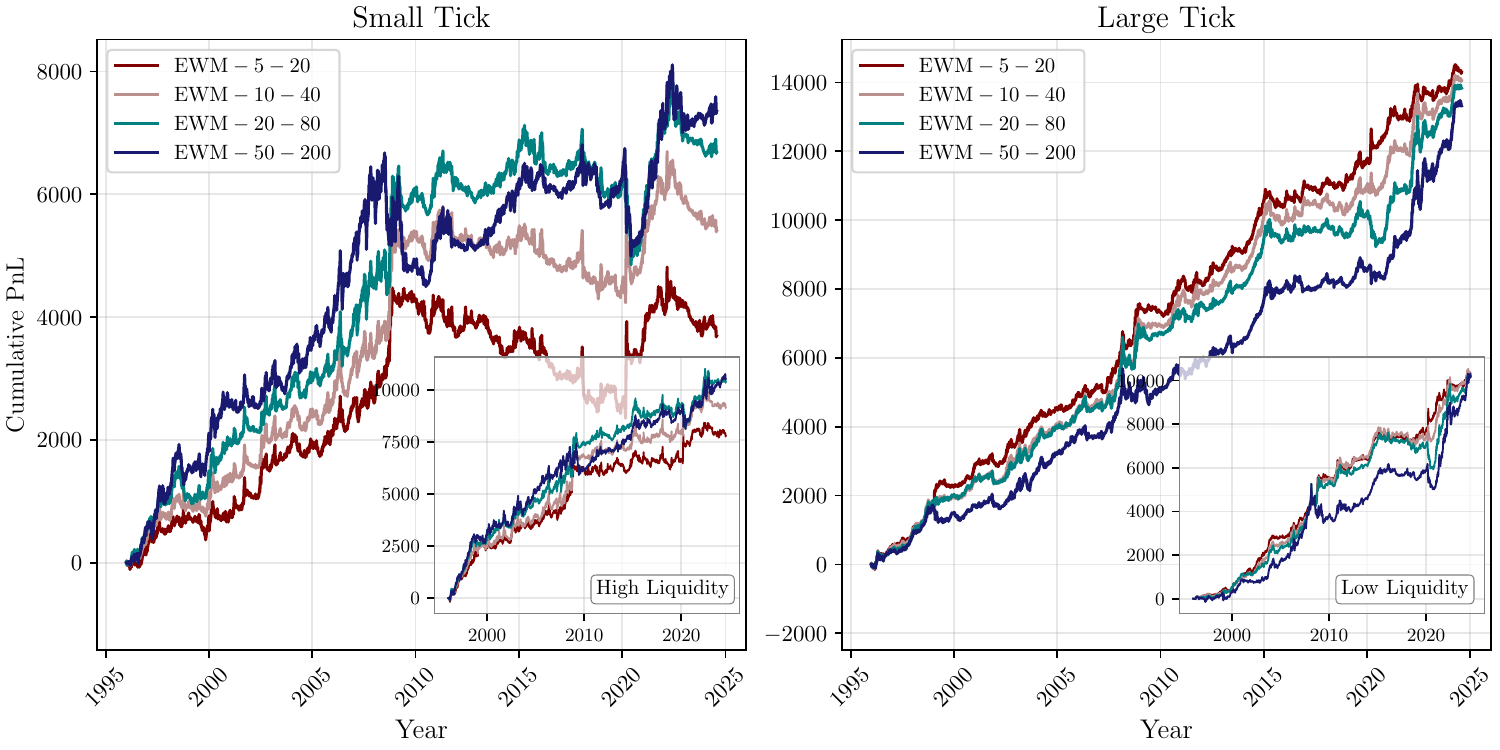}
\caption{Cumulative trend PnL conditioned on the monthly tick-size
tier. Left: small-tick contracts. Right: large-tick contracts.
Equal-risk portfolio, no liquidity weighting
(Eq.~\eqref{eq:portflio_wo_liq_weighting}). Insets: same, but
filtered by equally-sized liquidity tiers instead of tick-size
tiers.}
\label{fig:pnl_ticksize}
\end{figure}

Fig.~\ref{fig:pnl_ticksize} shows the cumulative PnL of the
equal-risk trend portfolio (Eq.~\eqref{eq:portflio_wo_liq_weighting},
without liquidity weighting), causally conditioned on the monthly
tick-size tier and disaggregated by signal horizon. The dichotomy
is stark: short-trend PnL has \emph{completely} degraded for
small-tick contracts after 2008, uniformly across signal horizons
EWM-$\tau$-$4\tau$, while the cumulative PnL of large-tick contracts
is virtually unaffected by the break and continues to accrue at
roughly the pre-2009 rate, even at the highest frequency. The result is preserved when the tick-size tiering is performed \textit{within} each asset class rather than across the full universe (Appendix~\ref{app:tick_tier_PnL_byAC}), suggesting that the dependence of fast-trend degradation on volatility-normalised tick size is approximately monotonic rather than driven by a single threshold.

\begin{figure}[H]
  \centering
  \includegraphics[width=0.7\linewidth]{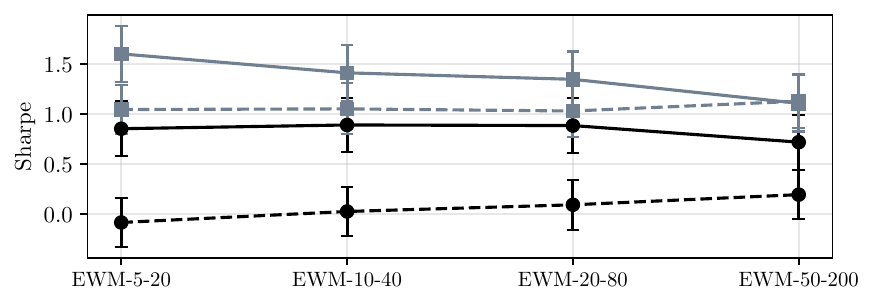}\\
\includegraphics[width=0.7\textwidth]{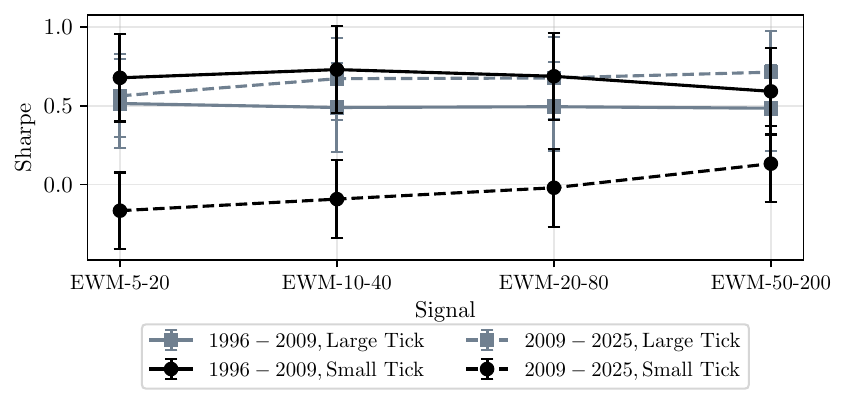}
  \captionof{figure}{Sharpe ratios by tick-size tier and signal
  horizon, pre- and post-break. Top: equal-risk portfolio. Bottom:
  liquidity-weighted portfolio. Errors are bootstrap standard
  errors.}
  \label{fig:sharpe_ticksize}
\end{figure}

Fig.~\ref{fig:sharpe_ticksize} (top) quantifies the dichotomy.
Pre-break Sharpes cluster around 0.8 for ST and 1.4 for LT for the
equal-risk portfolio. Post-break, ST Sharpes collapse to essentially
zero (mildly negative for the fastest signals), while LT Sharpes
remain in the range 1.0--1.2. The relative degradation is $\sim$100\%
for ST and $\sim$0--30\% for LT. Fig.~\ref{fig:sharpe_ticksize}
(bottom) repeats the analysis for the liquidity-weighted portfolio
(Eq.~\eqref{eq:portfolio_liqw}), from which qualitatively similar
insights may be inferred -- the finding holds with and without
liquidity weighting. Beyond that, the comparison also shows that
liquidity has a measurable but secondary effect on the PnL outcome.

\paragraph{Classifying by liquidity does not work.} A natural concern is that
liquidity might tier the cross-section equally well, since liquidity
and tick size are negatively correlated (log-log Pearson correlation
of $-0.35\pm0.08$; see Appendix~\ref{app:liquidity}). It is also
anecdotally known that trends still work on lower-liquidity
products. The result, however, is unambiguously negative: the
decomposition by liquidity tier (Fig.~\ref{fig:pnl_ticksize}, insets;
see also Fig.~\ref{fig:PnL_by_LiquidTier} in
Appendix~\ref{app:liquidity}) does not dichotomise as cleanly.
Liquidity-tiered PnLs are inconsistent across horizons: not all
fast signals on liquid contracts have collapsed, and there is no
clean ordering between the high- and low-liquidity sub-portfolios.
\emph{Tick size -- rather than liquidity, asset class, or
electronification -- is the cross-sectional variable that best
explains trend survival.} The asset-class heterogeneity documented
earlier is, on this reading, a downstream symptom of the clustering
of IDX and FXR in the small-tick tier and of YLD and most CMD in
the large-tick tier.

\paragraph{Within-contract confirmation.} Two further decompositions
confirm the framework. First, partitioning each contract's days by
volatility (per-contract per-year median, causal) shows that trends
are generally more profitable in periods of low volatility -- the
\textit{LeBaron effect}
\citep{lebaron1992serialCorrelvsVola, lebaron1992persistence_vola}.
Interestingly, low-volatility periods on small-tick contracts
continue to accrue PnL even after the overall PnL break: small-tick
products in low-volatility regimes behave similarly to large-tick
contracts -- exactly as expected if the operative variable is
$\Psi/\sigma$ rather than $\Psi$ alone. Second, partitioning by
daily return magnitude shows that the trend break is a phenomenon
of \emph{large-return} days only: the PnL contribution from
sub-1$\sigma$ days is essentially invariant to the post-2008
regime change across all horizons and tiers. The signal harvested
on small returns has not decayed; what has been eliminated is
specifically the trend follower's ability to profit from large
directional moves on small-tick contracts. Both decompositions are
reported in detail in Appendix~\ref{app:decompositions} and impose
constraints that the mechanism developed below naturally satisfies.

\section{HFT Liquidity, the Self-Fulfilling Loop, and the
Sparse-vs.-Dense Asymmetry}
\label{sec:mechanism}

We now turn to the mechanism. The organising idea is the
self-fulfilling loop introduced in Sec.~\ref{sec:intro}: trend
exists, and remains profitable, because trend trades push prices in
the direction of the signal, sustaining it for the next round of
trend followers to act on. The loop has two preconditions. First,
trend followers must be able to execute aggressively at a cost that
does not exceed the impact-mediated alpha they thereby create.
Second, the relationship between aggressive flow and price -- the
impact function -- must remain intact, so that flow continues to
translate into the price moves the signal needs.

We argue in this section that both preconditions held until roughly
2010 across the futures universe, and that since then both have
been compromised on small-tick contracts and largely preserved on
large-tick ones. The mechanism behind this asymmetry is the
post-crisis transition to HFT-dominated market making, which
operates in both tick-size tiers but has profoundly different
consequences depending on the geometry of the limit order book
(LOB). The empirical signature of the broken loop, we will see, is
twofold: a rotation of the LOB against trend execution
(Sec.~\ref{sec:mechanism_rotation}), and a flattening of the
contemporaneous return-versus-book-imbalance relationship that
quantifies the cost of providing depth in front of directional flow
(Sec.~\ref{sec:mechanism_impact}). Together, these two changes are
sufficient to explain why both the PnL and the signal itself have
collapsed on small-tick contracts.

\subsection{Liquidity provision against trend flow has rotated}
\label{sec:mechanism_rotation}

\begin{minipage}{0.39\textwidth}
We first revisit the order-flow diagnostics of
Sec.~\ref{sec:H1H2H3} but disaggregated by tick-size tier rather
than by asset class. The result is shown in
Fig.~\ref{fig:corr_stlt}. For slow signals, no material change in
the correlations with $I^\textup{book}$ (left) is visible in either
tier. For fast signals on \emph{large-tick} contracts, the
correlation
$\text{Corr}(\Delta\pi^{\text{CTA}}, I^{\text{book}})$ has changed
sharply from $\sim-3\%$ p.a.\ pre-2010 to $\sim+4\%$ p.a.\
thereafter. To wit: pre-2010, on a fast trend buy day
($\Delta\pi^{\text{CTA}}>0$), the negative correlation implies
$V^{\text{ask}}>V^{\text{bid}}$ -- there was more resting volume
offered to trend followers than to mean-reverters; post-2010, the
positive correlation implies $V^{\text{ask}}<V^{\text{bid}}$ --
the liquidity in the book has rotated \emph{against} the trend
follower.

For \emph{small-tick} contracts, the correlation has been positive
throughout, but its post-2010 magnitude on fast signals has grown
materially: the small-tick LOB, never abundant in liquidity in the
direction of trend, has become even more shallow.
\end{minipage}%
\hfill
\begin{minipage}{0.59\textwidth}
    \centering
\includegraphics[width=\textwidth]{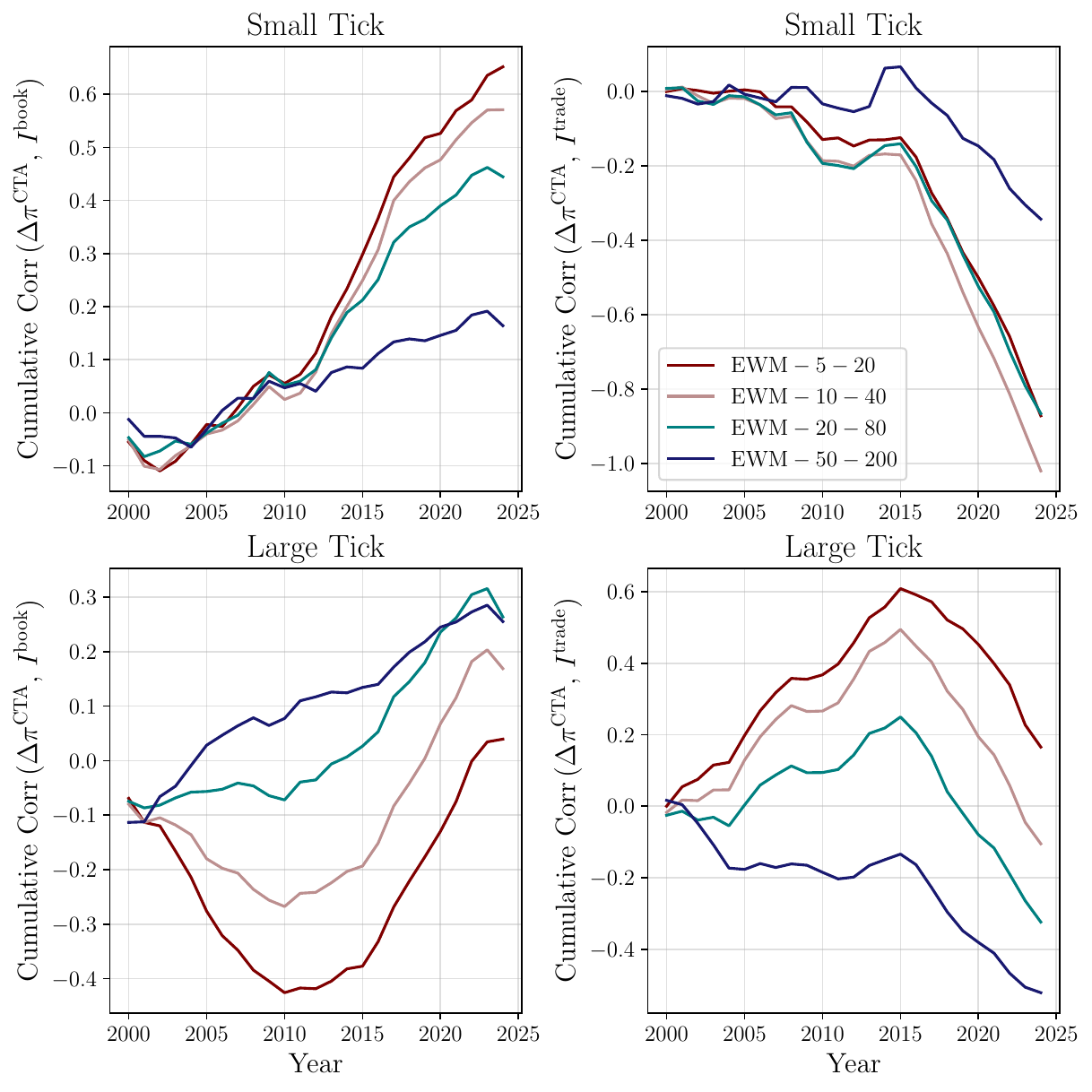}
\captionof{figure}{Cumulative annual-average correlation between
CTA trades (Eq.~\ref{eq:scaleup}, liquidity-weighted) and book
imbalance (left) / trade imbalance (right). Top: small-tick tier.
Bottom: large-tick tier.}
\label{fig:corr_stlt}
\end{minipage}\vspace{0.1cm}

The correlation with trade imbalance in
Fig.~\ref{fig:corr_stlt} (right column) also shows a clear regime
change, but only in 2015. This suggests either a change in the
execution strategy of trend followers or more market-order execution
from mean-reverters; in either case, this change is more a
consequence than a cause of the demise of short trends from 2010
onwards.

\subsection{The mechanism}
\label{sec:mechanism_mechanism}

The puzzle to be explained is therefore the following: post-2010,
the LOB has rotated against fast trend execution across
\emph{both} tick-size tiers, yet the PnL consequences are sharply
asymmetric -- innocuous for large ticks and devastating for small
ticks. We propose that the post-2010 break results from the
combination of two ingredients.

\paragraph{(i) A change in the liquidity-provision regime.} The
post-2008 period coincides with the substantial completion of the
transition from traditional bank-affiliated and proprietary-desk
market making, operating on inventory horizons of hours to days, to
high-frequency-trading market making on intraday flat-inventory
mandates with extremely tight spread-capture economics
\citep{kirilenko2017flashcrash, menkveld2013HFT,
chaboud2014riseAlgoTrading, hendershott2013algorithmic}. The HFT
market-making business model is structurally incompatible with
absorbing the predictable and persistent directional flow that
aggregate CTA trading generates. The empirical literature documents
that HFTs systematically \emph{withdraw} liquidity in front of large
institutional orders rather than supply it, and that execution costs
for such orders are correspondingly elevated
\citep{korajczyk2019HFT, vanKervel2019HFTvsInsti}. Whether the
relevant change is one of \emph{identity} (a new population of
providers, who never absorbed such flow) or \emph{behaviour}
(incumbents tightening risk preferences after the 2008 inventory
losses; see \citealt{adrian2010changing}) is immaterial for our
argument; the evidence is consistent with both, plausibly in
combination. The correlation changes reported in
Fig.~\ref{fig:corr_stlt} (left column) indicate that LOB liquidity
provision has rotated against trend followers, most probably as a
consequence of this evolution of market-making practices.

\paragraph{(ii) Structural differences between small- and large-tick
LOBs.} Independent of any particular generation of market makers,
small-tick LOBs are intrinsically \emph{sparse}: gaining priority
requires offering better prices, so expected gains are smaller
relative to adverse-selection costs; equilibrium posted depth is
therefore small, and gaps between filled price levels are common
\citep{bouchaud2018trades, dayri2015large}. Large-tick LOBs are
intrinsically \emph{dense}: the spread is meaningfully wider than
what is needed to compensate for adverse selection, the queueing
rent at the front of the queue is substantial, and resting volume
builds up at multiple deep levels.

\paragraph{Combined consequence: the loop holds on large ticks and
breaks on small ticks.} The HFT-driven rotation of the book operates
in both tiers (Fig.~\ref{fig:corr_stlt}), but its consequence is
asymmetric. In a sparse small-tick LOB, withdrawal removes the
residual depth that previously allowed fast trend execution to
proceed at reasonable cost -- which is precisely what the
self-fulfilling loop requires, since the CTA industry's flow comes
in large size. Trend followers would have had to ``walk the book''
to execute, incurring substantial transaction costs; this is the
short transitory period before they acknowledged that liquidity
previously available to them had significantly receded. The
consequence, in our scenario, is that trend followers structurally
disengaged from short-horizon trend in those products
\citep{korajczyk2019HFT}. Once they did so, the loop broke on its
input side: the impact-mediated reinforcement of nascent trends
disappeared along with the flow that produced it, and the small-tick
\emph{signal} -- not just the harvest -- decayed. This is exactly
the pattern documented in Sec.~\ref{sec:H1H2H3} under zero-lag
execution and corroborated by the within-contract decompositions
of Sec.~\ref{sec:ticksize}.

\begin{figure}[htbp]
    \centering
    \includegraphics[width=0.9\linewidth]{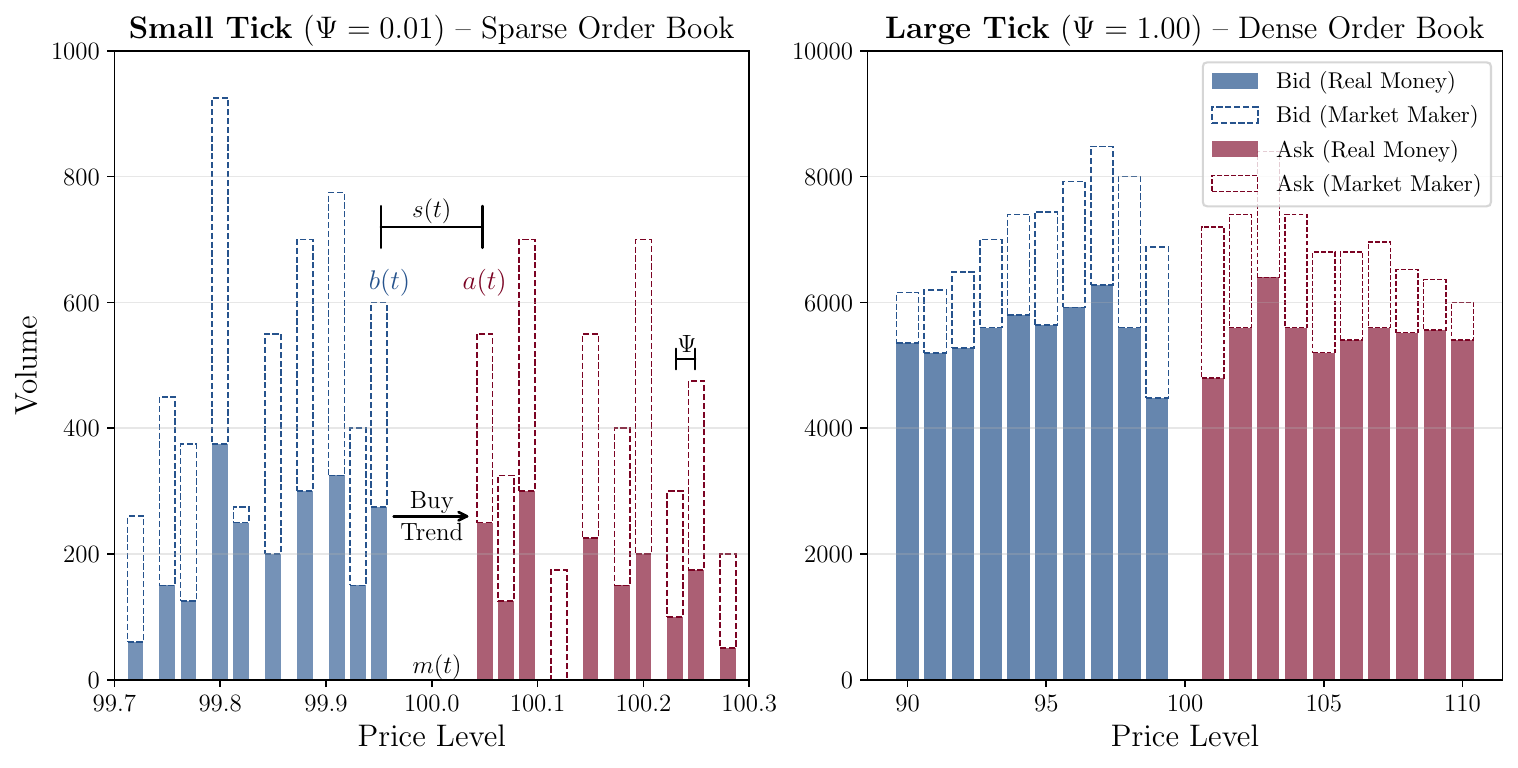}
    \caption{Limit order book sketches for a small-tick (left) and
    large-tick (right) asset. Dashed bars exemplify the liquidity
    provided by market makers, while solid bars subsume all other
    volume. The units $[\Psi]=\;$[price level] are arbitrary
    (futures) price units and can differ between contracts; $a(t)$
    and $b(t)$ are the best ask and bid prices, which can move in
    units of the tick $\Psi$; $s(t)=a(t)-b(t)$ is the spread, and
    $m(t)=\frac{a(t)+b(t)}{2}$ the mid-price. Note the difference
    in order of magnitude on the y-axes.}
    \label{fig:LOB_sketch_STLT_LiqWithdraw}
\end{figure}

By contrast, on dense large-tick LOBs, residual depth at the best
quotes and at deeper levels remains sufficient for execution to
proceed largely unperturbed compared to the pre-2010 period. Trend
followers continue to trade aggressively, the impact-mediated
reinforcement continues to operate, and both the signal and the PnL
remain intact. Fig.~\ref{fig:LOB_sketch_STLT_LiqWithdraw} sketches
the sparse small-tick and dense large-tick LOBs respectively, and
illustrates the asymmetric effect of liquidity withdrawal under a
buy-trend.

This account is also consistent with the within-contract
decompositions of Sec.~\ref{sec:ticksize} and
Appendix~\ref{app:decompositions}. The volatility decomposition is
mechanically predicted: in low-volatility regimes, $\Psi/\sigma$
rises and small-tick contracts behave microstructurally like
larger-tick ones, so the loop continues to operate. The
return-magnitude decomposition is also mechanically predicted:
liquidity withdrawal is triggered by the size of detected
directional flow, and the adverse-selection cost of providing
liquidity is small for weak trend signals -- hence depth does not
recede significantly and small-move trend PnLs are mostly
unaffected.

\subsection{The role of price impact: the loop, made explicit}
\label{sec:mechanism_impact}

\begin{figure}[h]
\centering
\includegraphics[width=0.49\textwidth]{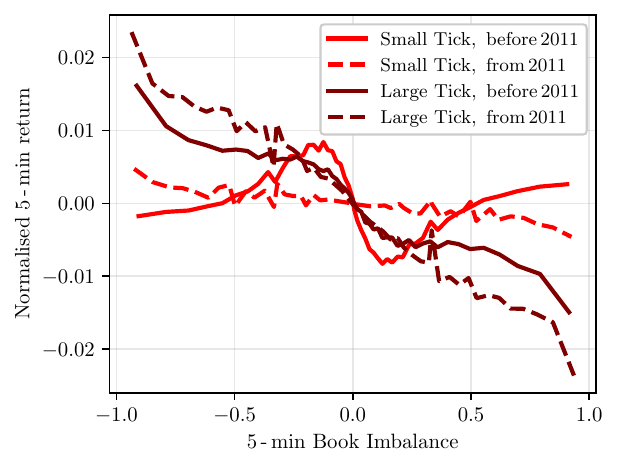}
\includegraphics[width=0.49\textwidth]{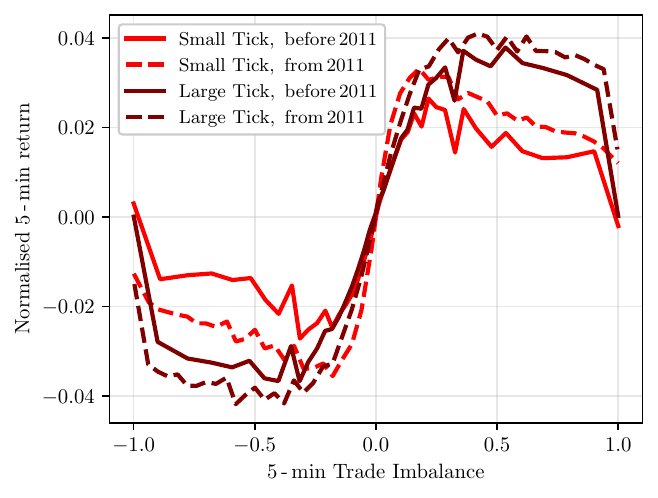}
\caption{Normalised 5-minute returns as a function of contemporaneous
book imbalance (left) and trade imbalance (right), for small-tick
(red) and large-tick (maroon) contracts, before (solid) and from
(dashed) 2011. The x-axis is quantile-binned.}
\label{fig:impact}
\end{figure}

Sec.~\ref{sec:mechanism_mechanism} argued that the degradation of
not only the trend PnL but also the trend \emph{signal} on small-tick
contracts is a consequence of trend followers' retreat from those
products. This is more than a casual claim: it is a direct
prediction of the self-fulfilling loop articulated in
Sec.~\ref{sec:intro}. \cite{deLong1990positivefeedback} showed that
trend signals are subject to positive feedback, so that more trend
trading should reinforce, rather than crowd out, the signal.
\cite{van2024ponzi} have recently made this scenario very precise
in the ETF context, quantifying the role of price impact; see also
\cite{patzelt2018universal} and Fig.~\ref{fig:impact} (right) for
microstructural evidence. The right panel shows the contemporaneous
relationship of normalised five-minute returns to the equal-bin
trade imbalance: under a buy-trend for which statistically
$V^\textup{BIV}>V^\textup{SIV}$ and hence $I^\textup{trade}>0$,
the return is on average positive, mechanically reinforcing the
trend signal. This is the loop, observable.

The implication for our story is straightforward. If trend
followers stopped trading aggressively on small-tick contracts, the
positive impact channel that fed back into the signal weakened
correspondingly: weak nascent trends could no longer be amplified
into full-fledged ones, and the signal's autocorrelation -- not
just its harvest -- decayed. On large-tick contracts, where
aggressive execution remained viable, the loop continued to
operate and the signal continued to exist.

Fig.~\ref{fig:impact} (left) sharpens the picture by showing the
relationship between the five-minute normalised return and its
equal-bin \emph{book} imbalance. It reveals striking differences
between small ticks and large ticks, and -- more importantly --
between small ticks before and from 2011, while the
return-versus-trade-imbalance relation (right panel) is stable
over time and qualitatively similar across tiers.

What do the small-tick curves tell us? Before 2011, there is clear
adverse selection: liquidity imbalance between bid and ask is
negatively correlated with the return. Limit orders at the bid are
likely to be executed at a higher price than the bar's close, and
vice versa, so liquidity providers face net inventory losses. The
average adverse selection reaches a maximum at intermediate book
imbalances ($I^\textup{book}\approx 0.1$; see
Fig.~\ref{fig:impact} (left, red)). Since 2011, however, this
effect has nearly entirely disappeared -- the correlation is close
to zero (dashed red line). This observation supports our hypothesis
that liquidity providers (in particular HFTs) in small-tick futures
have become largely immune to adverse selection, or at least
considerably more wary of it than in the past. The mechanical
signature of liquidity withdrawal is precisely this flattening:
depth that would previously have been ``run over'' by aggressive
trend flow no longer rests long enough to appear in the imbalance
statistics. The cost has not disappeared; it has migrated from the
market maker's inventory to the trend follower's slippage.

The correlation is persistently negative for large ticks, which is
expected to some extent: market-maker profits on large spreads can
remain positive even when partly eaten up by adverse selection.

\subsection{Limit orders are no remedy: missed opportunities and a
broken loop}
\label{sec:mechanism_alt_execution}

Can fast trend followers
circumvent the small-tick constraint by switching from market-order
to limit-order execution? We believe not, for two distinct but
mutually reinforcing reasons, both visible in the empirical
price-impact relationships of Fig.~\ref{fig:impact}.

\paragraph{(a) Trend limit orders are penalised by missed opportunities, not
adverse selection.} A trend follower posting a passive bid in order
to buy faces the risk of not being executed if their prediction is realized, 
leading to an increased execution price if the decision is not executed quickly.
The dominant cost of passive execution for a trend follower is therefore not adverse
selection in the usual sense (being filled by better-informed
counterparties before the price moves against them); it is the
opposite -- the \emph{missed-opportunity cost} of failing to be
filled while the price runs away in the direction of the signal.
The very fact that the signal is informative about the next move
guarantees that the follower's passive bid will be filled
preferentially in the bad states of the world (when the price drops
through it) and missed in the good ones (when the price runs away
upward). Limit-order execution therefore does not save costs; it
re-labels them, replacing visible slippage with invisible
opportunity loss. More generally, limit order execution is favoured in 
mean-reverting environments, and detrimental in trend following environments (see e.g. \cite{bouchaud2018trades}, chapter 17).

\paragraph{(b) Passive execution forfeits the self-fulfilling loop.}
The deeper problem is that even a costlessly executed limit order
would not solve the trend follower's predicament, because passive
trades do not feed the self-reinforcing impact channel that
sustains the signal in the first place. The trade-imbalance impact
of Fig.~\ref{fig:impact} (right) is generated by \emph{aggressive}
flow: it is the buyer-initiated volume that pushes prices up, not
the resting bid that sits and waits. A trend follower who replaces
market orders with limit orders ceases to contribute to
$I^\textup{trade}$ in the direction of the trend, and -- by exactly
the mechanism formalised in Sec.~\ref{sec:mechanism_impact} -- no
longer participates in the impact-mediated reinforcement of the
very price moves they are trying to exploit. The loop, on which
trend depends both for its profitability and for its existence, is
broken from the trend follower's side.

\paragraph{} Taken together, (a) and (b) leave no execution style
that side-steps the post-2008 microstructural change on small-tick
contracts. Aggressive execution is constrained by the thinness of
the post-withdrawal small-tick book; passive execution incurs
missed-opportunity costs that scale with the very informativeness
of the signal, and forfeits the impact channel that makes the
signal worth trading in the first place. This is, in our view, the
fundamental reason why the small-tick break has proven structural:
it is not an artefact of any particular implementation choice, but
a consequence of the relationship between aggressive flow and price
formation on which trend itself depends.

\section{Discussion and Conclusion}
\label{sec:conclusion}

The post-2008 decay of trend-following returns has been one of the
most discussed -- and least understood -- regularities in the
recent history of systematic investing (see e.g. \cite{lemperiere2014two, schmidhuber2021trend_reversion, quantica2025trend_capacity}). We have argued that, at
root, it is neither a flow phenomenon, nor a capacity phenomenon,
nor a consequence of electronification per se, but a microstructural
one. The volatility-normalised tick size cleanly partitions the
futures universe into a small-tick subset on which fast trend has
collapsed and a large-tick subset on which it has remained
essentially intact. No decomposition by asset class, liquidity, or
sectoral electronification date reproduces this dichotomy.

The interpretive frame we propose takes the self-fulfilling impact
loop -- signal $\to$ trade $\to$ impact $\to$ reinforced signal --
as the mechanism through which trend exists in the first place.
Trend followers do not merely harvest a pre-existing anomaly: their
aggressive directional flow, mediated by impact, sustains the very
price patterns they trade on. The loop has two preconditions: that
aggressive execution be feasible at reasonable cost, and that the
relationship between aggressive flow and price remain intact. Both
held until roughly 2010 across the futures universe; since then,
both have been compromised on small-tick contracts and preserved
on large-tick ones.

The mechanism behind this asymmetry is the post-2008 transition to
HFT-dominated market making, which replaced a generation of
inventory-tolerant liquidity providers with one whose business
model is structurally incompatible with absorbing predictable
directional flow. The resulting liquidity withdrawal in front of
CTA orders operates in both tick-size tiers, but its consequences
are asymmetric. On dense large-tick books, residual depth at the
best quotes and at deeper levels remains sufficient for execution
to proceed largely unperturbed; the loop continues to turn, and
both the signal and the PnL survive. On sparse small-tick books,
withdrawal removes the residual depth that previously supported
fast trend execution, forcing trend followers either to ``walk the
book'' aggressively or to retreat from the contract altogether. As
they retreated, the loop broke on its input side: the
impact-mediated reinforcement of nascent trends disappeared along
with the flow that produced it. The collapse of the small-tick
trend signal under zero-lag execution -- for which execution
costs is not an issue -- is the cleanest empirical fingerprint of this
broken loop. The within-contract decompositions by volatility and
by return magnitude confirm that the operative variable is
$\Psi/\sigma$ rather than $\Psi$ alone, and that the break activates
precisely on the large directional moves that would, under the old
regime, have triggered absorption by inventory-tolerant market
makers.

The contemporaneous return-versus-book-imbalance relationship makes
the same story visible at intraday frequency. Its pre-2011
small-tick negative steepness measures the cost of providing depth in front
of directional flow -- exactly the cost the previous generation of
market makers were absorbing. Its post-2011 flattening is the
mechanical signature of liquidity withdrawal: depth that would have
been run over no longer rests long enough to appear in the
imbalance statistics. 

Some practical implications follow. The break is structural; we
see no plausible scenario in which it reverts absent a
corresponding structural change in liquidity provision -- whether
through regulation, the entry of a new class of inventory-tolerant
intermediaries, or a re-tariffing of market-making rents. Switching
from aggressive to passive execution is not an option: passive
execution incurs opportunity costs and forfeits the
self-reinforcement channel that sustains the signal. Capacity
estimation for trend portfolios should accordingly be carried out
at the tick-size-tier level rather than at the asset-class level;
aggregate participation rates can be substantially below historical
bounds while still being binding within the small-tick subset.

Several questions remain open. Direct tests of the mechanism using
HFT inventory data around large CTA rebalancing events would sharpen
the causal claim and quantify the speed at which depth is withdrawn
and refilled. The possibility of recovering trend in microstructural
niches -- contracts below the HFT participation threshold, venues
with different market-maker mandates, or markets where tick-size
regimes have shifted -- is an empirical question we leave to future
work. So is the symmetric one: whether the same mechanism
disadvantages other systematic strategies that generate predictable
directional institutional flow, and whether, conversely, strategies
that trade against such flow have benefited from a complementary
tailwind.

A more fundamental open question concerns the relative weight of
the self-fulfilling impact loop among the mechanisms that generate
trend in the first place. The literature has long offered
alternatives that do not rely on impact-mediated reinforcement:
gradual incorporation of public news under behavioural
underreaction \citep{hong1999unified, daniel1998investor, bouchaud2019sticky};
slow, staggered execution by informed participants who correctly
anticipate a future event and accumulate positions ahead of it;
and the gradual diffusion of private information through a
heterogeneously-informed investor base
\citep{hong2000bad}. These mechanisms are not mutually
exclusive with the loop we emphasise; on the contrary, they
plausibly coexist, with the loop amplifying trends that have an
exogenous origin and the impact channel translating slowly arriving
information into prices. Our reading of the post-2008 evidence is
that the impact-loop component has been disproportionately
weakened on small-tick contracts -- which is consistent with the
near-total collapse of fast trends there -- while the
information-diffusion component, operating on slower horizons,
remains largely intact, consistent with the partial survival of
slow trends across the universe. Disentangling these contributions
quantitatively, ideally by exploiting cross-sectional or
cross-venue variation in the strength of the impact channel,
strikes us as a particularly worthwhile direction for future work. In fact, we do have direct empirical evidence some traders did anticipate large market moves and created trends accordingly, and that this possibility has decayed around approximately the same period, see Fig.~\ref{fig:PnL_STLT_byReturnMagnitude} of Appendix~\ref{app:decompositions}.  

Finally, our analysis suggests that the joint design of tick size
and market-maker obligations has first-order consequences for the
persistence of medium-frequency anomalies sustained, in part, by
self-fulfilling impact loops; this is a topic on which the
regulatory and academic literature have arguably underweighted the
perspective of the directional taker of liquidity.

\section*{Acknowledgements}
We thank Adam A. Majewski for facilitating the access to bar data and for fruitful discussions. Further, we are grateful to Société Générale Prime Services \& Clearing for providing the SG CTA Index data, and to Robert Almgren (Quantitative Brokers) for the electronification data and helpful discussions.

\bibliographystyle{abbrvnat}
\bibliography{biblio}

\appendix

\section{Data}
\label{app:data}

\paragraph{Daily data.} All signal construction, portfolio simulation, and tick-size analyses use daily settlement prices and (daily reported) ticks for the following 101 futures contracts, spanning the four sectors introduced in Sec.~\ref{sec:portfolio} (commodities, equity indices, currencies, and government bonds/yields):

\begin{quote}\small
10USNOTE, 10YCAN, 2USNOTES, 5USNOTES, AEX, AUD, AUS10YR, AUS3YR, BOBL, BRENT0, BUND, BUSDBRL0, BUXL, CAC, CANOLA0, CCRUDE\_OIL0, CD, CHF, CIRON\_ORE0, COPPER0, CORN0, CPALM\_OIL0, CPTA0, CRAPESEEDM0, CRAPESEEDO0, CRUDE0, CSOYBEAN\_OIL0, CSOYMEAL0, DAX, DJMINI, EUR, EUROSTOXX, FCATTLE0, FRA10YR, FTSE, FTSEJSE40, FTSE\_CHINA\_A50, FTSE\_TAIWAN, GASOIL0, GBP, GILTS, GOLD0, HANGSENG, HEATOIL0, HRW0, HSHARES, IBEX, INSE\_NIFTY, ITA10YR, ITA3YR, JGB, JPY, KOR10YR, KOSPI, KRW, KTB3YR, LCATTLE0, LCOCOA0, LCOFFEE0, LEANHOGS0, LSUGAR0, MBOVESPA, MIB, MNIKKEI, MSCIEMMINI, MSCIS, MXP, NATGAS0, NATGAS\_TTF0, NATGAS\_UK0, NCOCCOA0, NCOFFEE0, NCOTTON0, NFCOJ0, NIKKEI, NIKKEIS, NSDQMINI, NSUGAR0, NZD, PALLADIUM0, PLATINUM0, PRAPESEED0, PWHEAT0, RBGASOL0, RUSSELLMINI, SCHATZ, SILVER0, SMI, SOYBEANM0, SOYBEANO0, SOYBEANS0, SPI200, SPMID, SPMINI, SPTSE60, TBOND, TOPIX, UTBOND, UTNOTE, WHEAT0, WTICRUDE0.
\end{quote}

\paragraph{Intraday bar data.} All order-book and order-flow analyses (Sec.~\ref{sec:H3_orderflow} and Sec.~\ref{sec:mechanism}) use proprietary 5-minute bar data from Capital Fund Management (CFM). Bar data are available for a smaller subset of contracts and over a shorter history than the daily series. We restrict the bar-data universe to contracts for which coverage begins no later than 2004, ensuring at least five years of pre-break data. This yields the following 53 contracts:
\begin{quote}\small
10USNOTE, 10YCAN, 2USNOTES, 5USNOTES, AUD, AUS3YR, BOBL, BRENT0, CD, CHF, COPPER0, CORN0, CRUDE0, DJMINI, EUR, FCATTLE0, FTSE, GASOIL0, GBP, GILTS, GOLD0, HANGSENG, HEATOIL0, HSHARES, JPY, KOSPI, KRW, KTB3YR, LCATTLE0, LCOCOA0, LCOFFEE0, LSUGAR0, MIB, MNIKKEI, MSCIS, MXP, NATGAS0, NCOCCOA0, NCOFFEE0, NCOTTON0, NIKKEI, NSDQMINI, NZD, RBGASOL0, RUSSELLMINI, SCHATZ, SILVER0, SPMINI, SPTSE60, TBOND, TOPIX, WHEAT0, WTICRUDE0.
\end{quote}

Each 5-minute bar reports, per contract, the total traded volume, the buyer- and seller-initiated volumes (cumulative over the interval), and end-of-interval snapshots of the mid-price and of the resting volumes at the best bid and best ask.

\section{Capacity Drag Calculation}
\label{app:capacity}
In this appendix, it is roughly quantified that the execution-cost drag implied by the square-root impact law is rather insufficient to explain the observed PnL collapse.

Assuming, on any given day, that the CTA industry trades a total quantity $Q$ over a time interval [0, $T$], starting from a initial position $q(0)=0$ and finishing with an aggregate position $Q=q(T)= \int_0^T \dot{q}(t) \mathrm{d}t$, where $\dot{q}$ are the trades (here in continuous time), then under the square-root impact model as in \cite{toth2011impact}, the total price change due to the CTAs should be
\begin{equation}
    p(T) = p(0) + \left( \frac{1}{V_\textbf{d}} \int_0^T \dot{q}(t) \,\mathrm{d}t \right)^{1/2} \sigma_\textup{d} Y + \text{noise},
\end{equation}
where $V_\textup{d}$ is the daily traded volume, $\sigma_\textup{d}$ is the daily volatility as a percentage of price (and thus in units of price: $[\sigma_\textup{d}]=[p]=\$$), and $Y\approx 1$ is an empirical constant. The noise process is zero-mean.

The most basic execution schedule, following a constant trading rate, $\dot{q}=$const, implies $\dot{q}= Q/T$, such that $q(t)=Q \frac{t}{T}$. Under this model, the expected price impact reads
\begin{equation}
    \mathcal{I} (t) = \mathbb{E} \left[ p(t) - p(0) \right] = \left( \frac{Q}{V_\textup{d} \frac{t}{T}} \right)^{1/2} \sigma_\textup{d} Y.
\end{equation}
The associated slippage or the impact-associated cost due to trading the entire position is
\begin{align}
    S(Q; T) &= \mathbb{E} \left[ \int_0^T (p(t)-p(0)) \underbrace{\dot{q}(t)}_{=\frac{Q}{T}} \, \mathrm{d}t \right] \\
    &= \int_0^T \left( \frac{Q}{V_\textup{d}} \frac{t}{T} \right)^{1/2} \sigma_\textup{d} Y \frac{Q}{T} \, \mathrm{d}t = \frac{2}{3} \sigma_\textup{d} Y \frac{Q^{3/2}}{V_\textup{d}^{1/2}}  \equiv S(Q),
\end{align}
implying a unit trading cost of
\begin{equation}
    \Bar{S} = \frac{S}{Q} = \frac{2}{3} \sigma_\textup{d} Y \sqrt{\frac{Q}{V_\textup{d}}}
\end{equation}
Therefore, the slippage in terms of daily volatility, which is unit-less, is the following for a participation rate of $1\%=\frac{Q}{V}$
\begin{equation}
    \frac{\Bar{S}}{\sigma_\textup{d}} = \frac{2}{3} Y \sqrt{\frac{Q}{V_\textup{d}}} = 
    \frac{2}{3} 1 \sqrt{0.01} = \frac{2}{30}.
\end{equation}
The annualised CTA trading cost, assuming 252 trading days, consequently is
\begin{equation}
    \mathcal{C}_\textup{ann} = \frac{2}{30} \mathcal{R}_\textup{d} \cdot \text{turnover} \cdot 252
\end{equation}
with a turnover, $\frac{\langle |\Delta q| \rangle}{\langle |q| \rangle}$, that is the fraction of the total CTA daily risk that is typically traded daily. This figure is around 9\% for the EWM-5-20 signal defined in Sec.~\ref{sec:TrendSignal_Definition}. $\mathcal{R}_\textup{d}$ is the daily CTA risk. With an annualised CTA risk of $\mathcal{R}_\textup{ann} = \sqrt{252} \mathcal{R}_\textup{d}$, the reduction in Sharpe ratio expected from this simple calculation with a participation rate of 1\% is
\begin{equation}
    \Delta \, \text{Sharpe} = - \frac{\mathcal{C}_\textup{ann}}{\sqrt{252} \mathcal{R}_\textup{d}} = - \frac{2}{30} 9\% \sqrt{252} \approx - 0.1. 
\end{equation}
The implied drag on Sharpe is therefore $\sim 0.1$ at the observed industry participation rate of approximately $1\%$. This is non-trivial -- it is roughly one-eighth of the pre-2009 trend Sharpe -- but is too small, even when allowing for model uncertainty of a factor of three to four, to explain the observed collapse from $\sim 0.7$ to $\sim 0$, even though the above should be treated rather as an order-of-magnitude calculation than a precise estimation.

\section{Volume-clock Decomposition}
\label{app:volumeclock}

The cumulative correlations of Sec.~\ref{sec:H3_orderflow} aggregate over the full trading day and may therefore be diluted if the flow they aim to detect is localised intra-day. Industry lore suggests that CTAs execute the bulk of their daily volume in the opening hours of the underlying cash markets; during that window their trades would be strongly correlated with the prevailing order-flow state, while during the remaining hours the correlation would be significantly lower. The daily-aggregate correlation then reflects, at best, a weighted average of a concentrated within-window signal and a long stretch of noise, mechanically underestimating the intra-day effect.

To probe any such structure, a \textit{volume-clock} is constructed. Because liquidity in futures is highly concentrated around the open of the underlying cash markets, a time-clock partition would assign vastly  different volumes -- and, therefore, vastly different signal-to-noise ratios -- to intervals of equal duration. A volume-clock partition, on the contrary, assigns equal numbers of trades to each segment, making the corresponding correlations directly comparable across bins.

\begin{figure}[ht!]
    \centering
    \includegraphics[width=\textwidth]{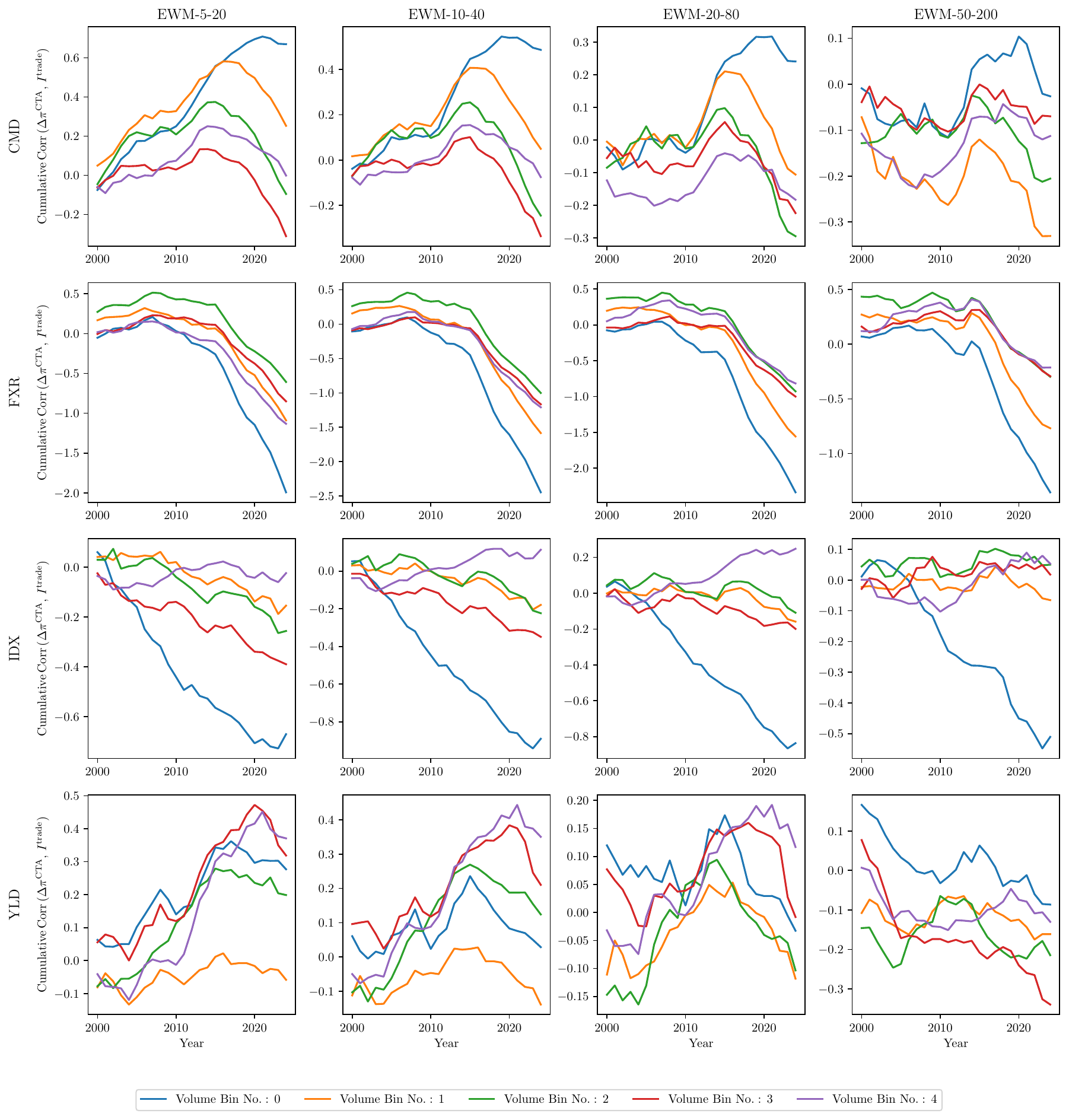}
    \caption{Cumulative correlations between daily CTA trades and the
    volume-weighted \textit{trade} imbalance computed within $20\%$-of-volume
    intra-day segments, by asset class (rows: CMD, FXR, IDX, YLD) and signal
    time scale (columns: EWM-$\tau$-$4\tau$ with $\tau \in \{5, 10, 20, 50\}$).
    Volume bin $0$ (blue) contains the first $20\%$ of daily volume.}
    \label{fig:vclock_trade}
\end{figure}

\begin{figure}[ht!]
    \centering
    \includegraphics[width=\textwidth]{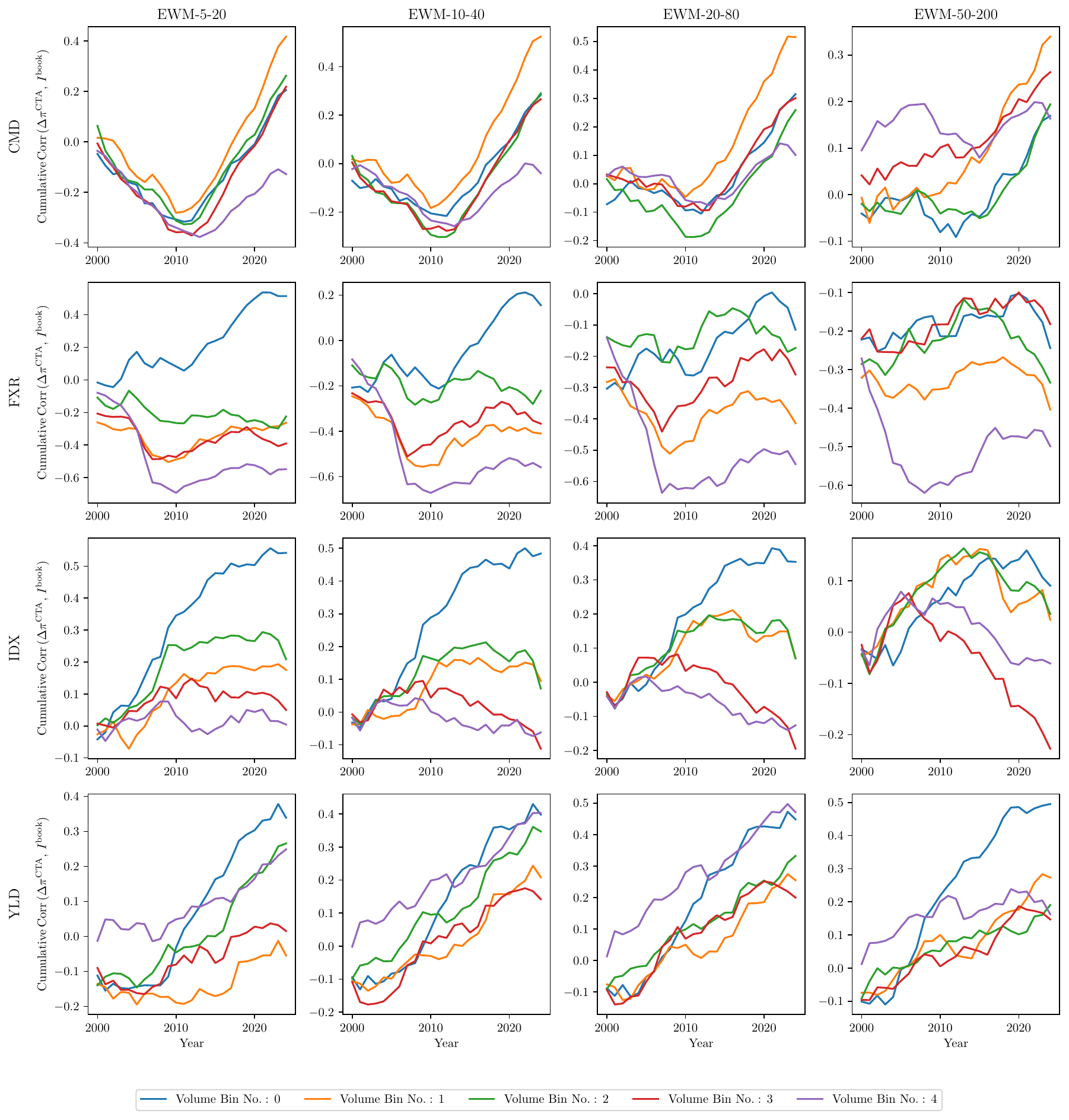}
    \caption{Same as Fig.~\ref{fig:vclock_trade} but for the volume-weighted
    \textit{book} imbalance.}
    \label{fig:vclock_book}
\end{figure}

\paragraph{Methodology.} For each product-day -- defined according to the local trading day of the underlying cash market -- we partition the 5-minute bars into five segments of equal traded volume, so that segment $0$ contains the first $20\%$ of daily volume, segment $1$ the next $20\%$, and so on. Within each segment, the book and trade imbalances of Eq.~\eqref{eq:imbalances} are recomputed using the segment's own volume as the normalisation rather than the daily one. The cumulative correlations between the daily CTA trades $\Delta\pi^{\text{CTA},\tau}$ and these segment-imbalances are then calculated as in Sec.~\ref{sec:H3_orderflow}, separately for each of the four asset classes and for four signal time scales $\tau \in \{5, 10, 20, 50\}$. Figs.~\ref{fig:vclock_trade} and \ref{fig:vclock_book} report the results.

\paragraph{Findings.} Two features stand out.

\textit{First, the asset-class heterogeneity persists.} For neither
imbalance does a single consistent picture emerge across CMD, FXR, IDX,
and YLD. The volume-clock decomposition does not resolve the cross-sectional inconsistency identified in Sec.~\ref{sec:H3_orderflow}; if anything, it reinforces it. This is, by itself, an important negative result: even the temporally most localised version of the H3 diagnostic fails to map the asset-class pattern of the order-flow regime change onto the asset-class pattern of PnL degradation.

\textit{Second, volume bin $0$ is an intra-day outlier systematically.}
This is the most consistent feature across asset classes -- though
its direction varies -- and is most conspicuous in the trade-imbalance
decomposition (Fig.~\ref{fig:vclock_trade}):

\begin{itemize}
\item For CMD, segment $0$ reaches the highest cumulative correlation in
the pre-2016 regime ($\sim +0.4$ for EWM-5-20) and is the only
segment that does not change sign afterwards.
\item For FXR, segment $0$ stands out from 2009 onwards, with the largest
(negative) cumulative correlation across signal speeds; segment 0 already
turns negative around 2009, while the other segments only join from 2016.
\item For IDX, segment $0$ is the most negative, reaching $\sim -0.7$ for
EWM-5-20, while other segments cluster at less negative or near-zero
values.
\item For YLD, segment $0$ does not stand out particularly.
\end{itemize}

The book-imbalance decomposition (Fig.~\ref{fig:vclock_book}) shows
qualitatively the same pattern: segment $0$ is systematically the outlier
-- most positive for FXR and IDX fast signals, in line with leaders for
YLD, and clustered with mid-segments for CMD.

\paragraph{Comparison with the daily aggregates.} The segment-wise
correlations in Figs.~\ref{fig:vclock_trade} and \ref{fig:vclock_book} are broadly comparable in magnitude to the daily-average per-asset-class correlations of Fig.~\ref{fig:corr_perAC}. The volume-clock decomposition therefore does not reveal an intra-day-localised effect that the daily aggregation had been underestimating. What it does reveal is intra-day structure -- specifically, that the first $20\%$ of daily volume behaves systematically differently from the remaining four segments -- which is masked in the daily average. In that sense the segmentation refines the diagnostic without amplifying it.

\paragraph{Interpretation.} Two readings are consistent with the extremity of segment $0$.

\textit{(a) CTA flow is intraday-concentrated.} If CTAs execute a disproportionate share of their daily volume within segment $0$ -- as industry lore suggests -- the correlation of their daily trades with the segment-$0$ imbalance is mechanically larger in magnitude than with segments where CTA activity is sparser. The sign of that correlation can in principle be sector-specific because the structure of liquidity provision at the open is itself asset-class-specific.

\textit{(b) Segment $0$ is structurally special.} The first $20\%$ of daily volume generally subsumes the overnight carry-over period and the high-volume opening of the underlying cash market -- periods of elevated volatility, liquidity fragmentation, and news incorporation. The order-flow statistics of segment $0$ can therefore differ from those of segments $1$--$4$ for reasons that have nothing to do with CTAs: any daily-aggregate flow would correlate differently with the segment-$0$ imbalance than with the others.

These readings are not mutually exclusive, and the data cannot cleanly attribute the bin-$0$ effect to one or the other. A decisive test of (a) would require knowledge of the actual intra-day distribution of CTA execution volume, which we do not have. What we can say is that the volume-clock result is consistent with the folklore of morning-concentrated CTA execution and constitutes -- to our knowledge -- the first systematic empirical support for it in the academic literature on futures (a similar concentration has been previously established for institutional equity flow by \citet{heston2010intraday} and \citet{bogousslavsky2016infrequent}).

\section{Within-contract Decompositions}
\label{app:decompositions}

For the tick size tiering that builds the groundwork in Sec.~\ref{sec:ticksize} is volatility-dependent, it is a natural concern whether the relationship between tick size and trend PnL goes yet deeper. Even within single contracts periods of high and low volatility -- impacting the liquidity-tier ranking through the denominator -- may correspond to different microstructural regimes, and may even have different PnL responses. This is was already alluded to in the final subsection of Sec.~\ref{label:ticksize} but is analysed here further. Beyond the volatility-decomposition, a second decomposition of the PnL into large- and small-return days is investigated, which will also constraint any mechanism attempting to explain trend performance decay.

\subsection{Decomposition by Volatility Regime}
\begin{figure}[b!]
\centering
\includegraphics[width=\linewidth]{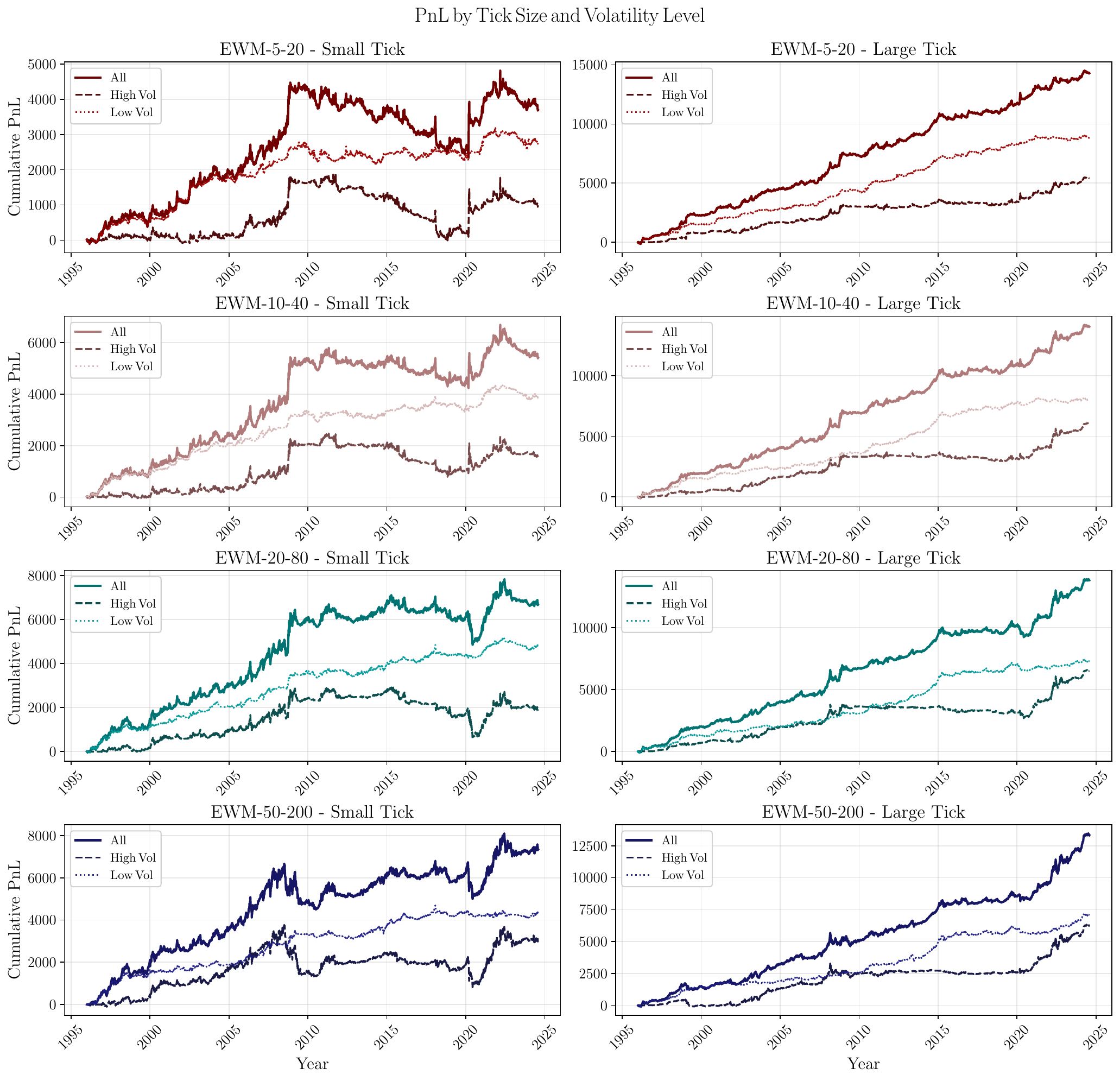}
\caption{Cumulative trend PnL (non-liquidity-weighted portfolio, Eq.~\eqref{eq:portflio_wo_liq_weighting}) decomposed into contributions from high-volatility and low-volatility days. Rows: signal horizon EWM-$\tau$-$4\tau$. Left column: small-tick tier. Right column: large-tick tier. The volatility regime is defined per-contract per-year using the median realised volatility, computed causally.}
\label{fig:PnL_STLT_byVol}
\end{figure}

For each contract $i$, a backward-looking (causal) rolling median $\sigma_{i}^\textup{med} (t)$ is computed on the last trading year (252 trading days). Based on this median, the trading days are partitioned into two similarly sized groups: a high-volatility ($\sigma_i (t)> \sigma_i^\textup{med} (t-1)$) section and a low-volatility section, the complementary case. The cumulative trend PnL is then computed on each sub-sample separately. The results for different trend signals are depicted in Fig.~\ref{fig:PnL_STLT_byVol}.

Several observations strike:
\begin{enumerate}
    \item Across all groups, low-volatility days contribute a disproportionate amount to the gains In large tick contracts the vast majority of PnL comes from periods of low volatility across all signal horizons. For small ticks the same holds true pre-2009, and even after the PnL break is less bad in low-volatility regimes.

    This observation is a strong cross-secctionla confirmation of the \textit{LeBaron effect} \citep{lebaron1992serialCorrelvsVola, lebaron1992persistence_vola}, which states that many systematic trading rules, including trend, perform better in low-volatility regimes. For some cross-sections, especially large ticks, and also small ticks, this effect seems to even have strengthened post-break as the distance between the low and high volatiltiy curves have widened.

    \item While the all-days PnL and even more so the high-volatility PnL has collapsed are even gone negative for small ticks, the low-volatility PnL continues to accrue, even if at a lower rate. Therfore, the post-2008 PnL collapse on small tick contracts can be attributed to high-volatility days.

    \item Small tick contracts behave qualitatively more like large tick products in periods of low-volatility. This observation may be interpreted as a confirmation of the effectiveness of the cross-sectional division based on $\Psi/\sigma_i$ (instead of $\sigma_i$ alone). Low-volatility periods correspond to a higher ratio, and on those days the LOBs of usually small tick products realtively resemble those of large tick products more closely in the microstructural sense and. The small tick trend PnL on those days also resembles the large tick one.
\end{enumerate}

\subsection{Decomposition by Return Magnitude}
The previous decomposition by volatility level hints at when the trend gains are collected. The next composition based on return magnitude shows from what kind of price changes trend gains are collected from.
A product $i$ is classified as \textit{large-return} when its current (volatility-normalised) absolute return is lower than the 1$\sigma$-threshold calculated causally based on the last 252 trading days, i.e. when $|r_i (t)|>1\sigma_i^\textup{ann} (t-1)$ and \textit{small-return} in the complementary case.

\begin{figure}[H]
\centering
\includegraphics[width=\linewidth]{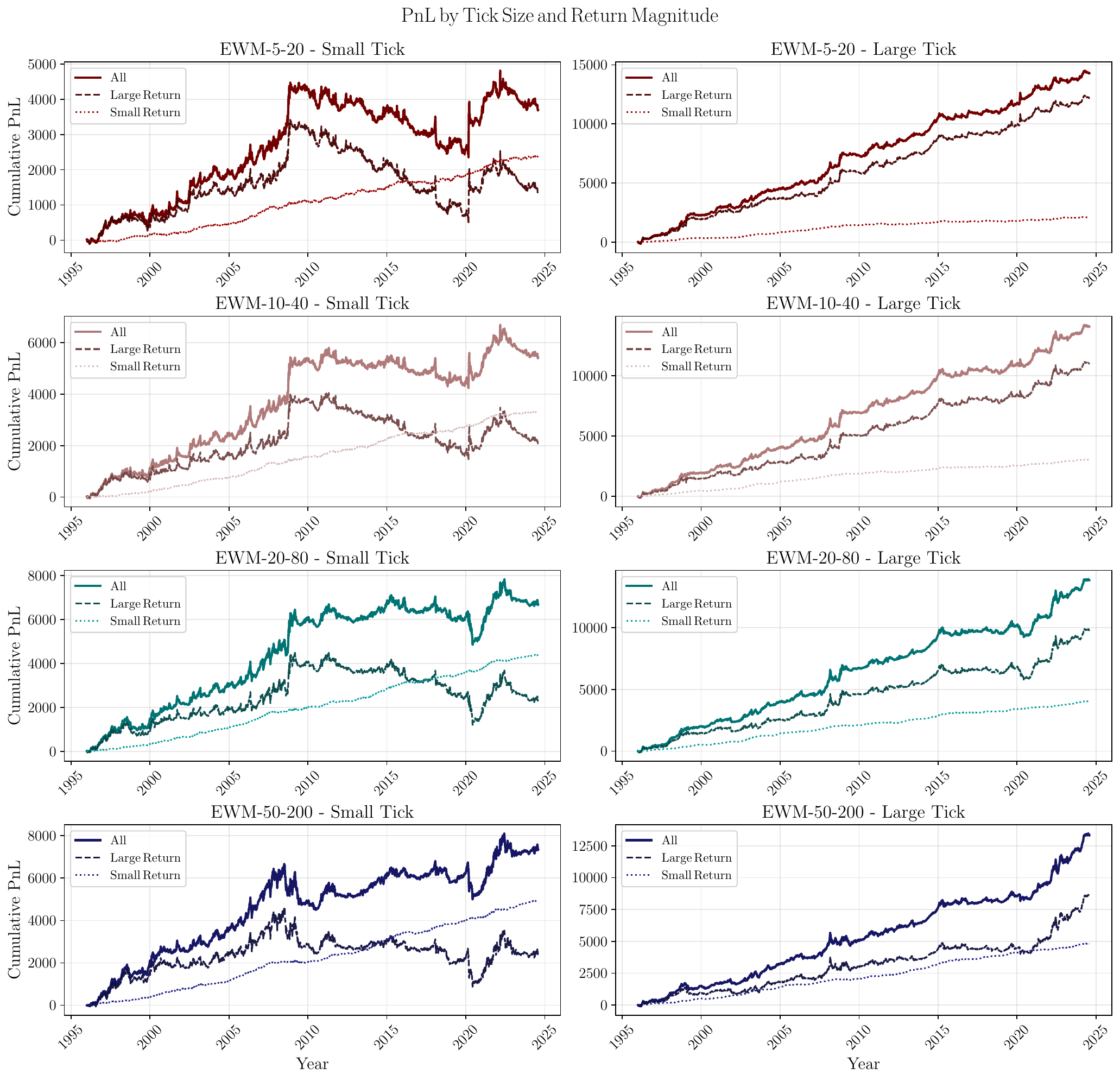}
\caption{Cumulative trend PnL (non-liquidity-weighted portfolio, Eq.~\eqref{eq:portflio_wo_liq_weighting}) decomposed into contributions from high-volatility and low-volatility days. Rows: signal horizon EWM-$\tau$-$4\tau$. Left column: small-tick tier. Right column: large-tick tier. The volatility regime is defined per-contract per-year using the median realised volatility, computed causally.}
\label{fig:PnL_STLT_byReturnMagnitude}
\end{figure}

Fig.~\ref{fig:PnL_STLT_byReturnMagnitude} reports the result, which is genuinely striking. The small-return-day PnL contribution is utterly \textit{invariant} to the trend break for all signals and for both tick size tiers. The small-return-PnL accrues at a nearly constant rate for the entire observation period 1995-2025, with no kink at the 2008/9 PnL break point, no acceleration during the 2008, 2014, or 2020 directional period, and no downturn after. In stark contrast stands the large-return-PnL that is responsible for basically all of the aberrations in the all-day PnL.

This allows for two important take-aways:
\begin{enumerate}
    \item The trend break is a phenomenon of large-return days, not of the underlying signal.
    \item The post-2008 regime shift has selectively eliminated trend followers' ability to profit from large price swings -- exactly the moves that previously allowed for the straegy's profitability.

    This imposes a stronger empirical constraint on the explanatory mechanism than the all-day PnL would alone: whatever the mechanism, it must selectively impact trend followers on large-return days, and must leave the strategy's performance under small returns unaffected.
\end{enumerate}

Fig.~\ref{fig:PnL_STLT_byReturnMagnitude} reports exactly that cross-sectional pattern.

Both decompositions therefore confirm the methodology deployed and consequences drawn:  the volatility decomposition confirms $\Psi_i/\sigma_i$ as the relevant microstructural variable, while the one by return magnitude highlights how trends stopped being predictive of large returns, thus significantly chipping away at the overall trend profitability.

\section{Liquidity Tier Decomposition}
\label{app:liquidity}

\begin{figure}[htbp]
    \centering
    \includegraphics[width=.6\textwidth]{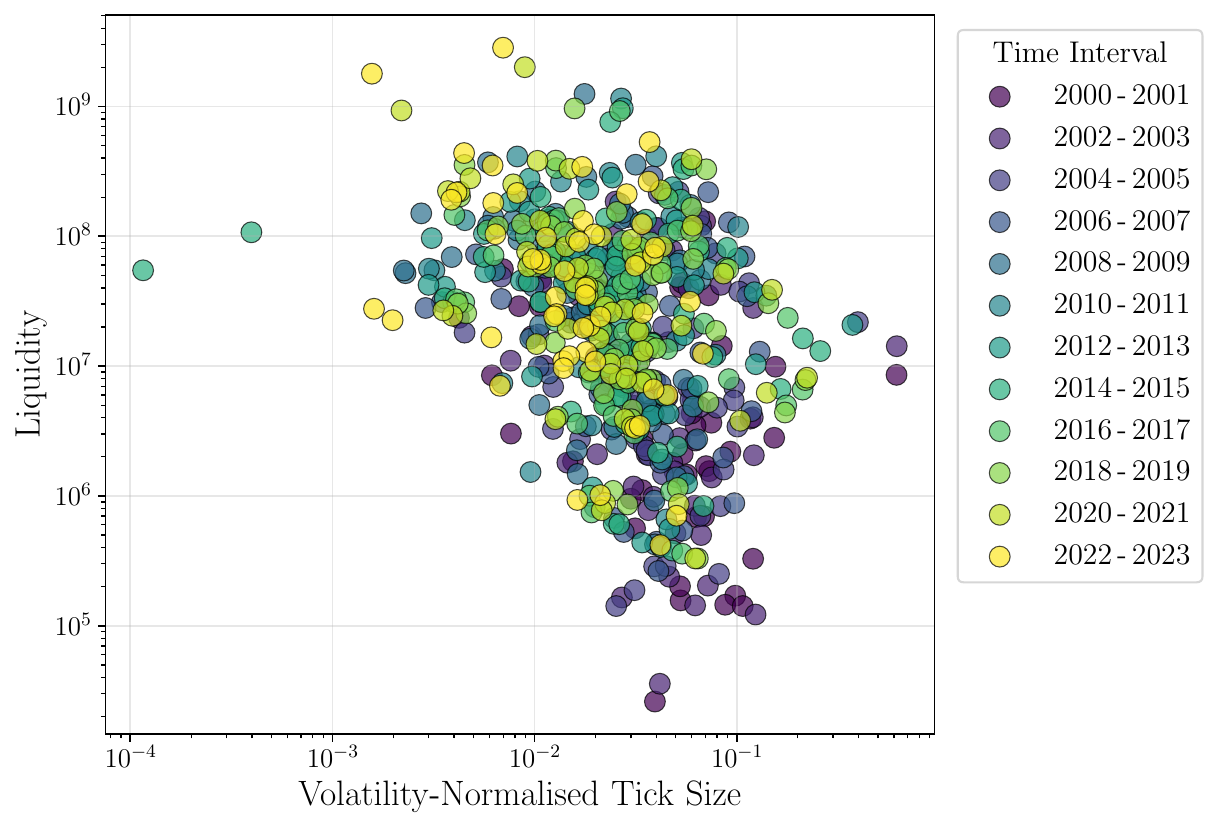}
    \caption{Liquidity vs. (volatility-normalised) tick size. The points are averaged in 2-year intervals. Axes are in log-scale; colours encode the progression in time.}
    \label{fig:Liq_vs_Ticksize}
\end{figure}

Fig.~\ref{fig:Liq_vs_Ticksize} graphically confirms the negative correlation between liquidity and tick size, because of which a trend analysis based on liquidity instead of tick size was investigated.

From the figure three aspects may be inferred: First, liquidity and volatility-normalised ticksize are anti-correlated with a Pearson-correlation of -$(35 \pm 8)$\% in log-log-scale. This means that more liquid contracts tend to have smaller volatility-normalised tick sizes. Second, (volatility-normalised) tick size has decreased over the years. Third, overall liquidity has increased markedly during the considered time span. The last two observations are known facts.

\begin{figure}[htbp]
    \centering
    \includegraphics[width=0.88\linewidth]{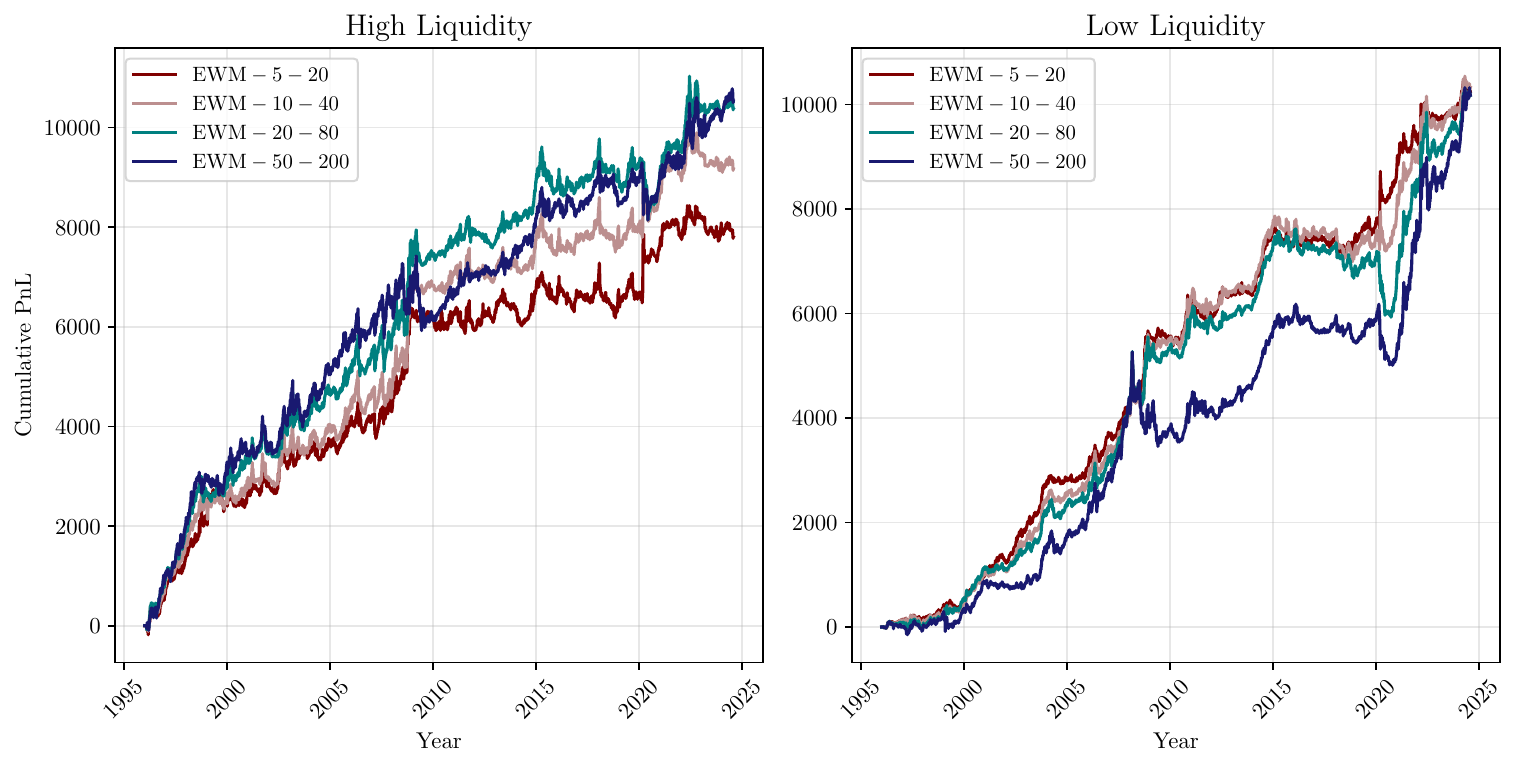}
    \caption{Cumulative PnL for different trend portfolios (not liquidity-weighted) and separated into two equal-sized liquidity tier. Tiering analogous to the procedure in Sec.~\ref{sec:ticksize} for the tick size.}
    \label{fig:PnL_by_LiquidTier}
\end{figure}

Fig.~\ref{fig:PnL_by_LiquidTier} shows decisively that a partition into high and low liquidity products does not explain trend degradation as well as (volatility-normalised) tick size in Sec.~\ref{sec:ticksize}, Fig.~\ref{fig:pnl_ticksize}. One caveat is that all 101 contracts in our universe are relatively liquid by construction: less standard, lower-liquidity products that sit below the HFT market-makers' radar may well retain trend profitability. Such products, however, tend to fall in the large-tick category, so their inclusion would, if anything, reinforce rather than contradict the mechanism proposed in Sec.\ref{sec:mechanism}.

\section{Within-Asset-Class Tick-Tiered Performance}
\label{app:tick_tier_PnL_byAC}
The tick size tiering procedure of Sec.~\ref{sec:ticksize} is repeated \textit{within} each of the four asset classes, partitioning each sector's products into the 50\% with the smallest and the 50\% with the largest volatility-normalised tick size. A commodity classified as small-tick under this scheme therefore belongs to the lower half of the commodity cross-section, not necessarily to the lower half of the full universe.

\begin{figure}[h]
    \centering
    \includegraphics[width=\linewidth]{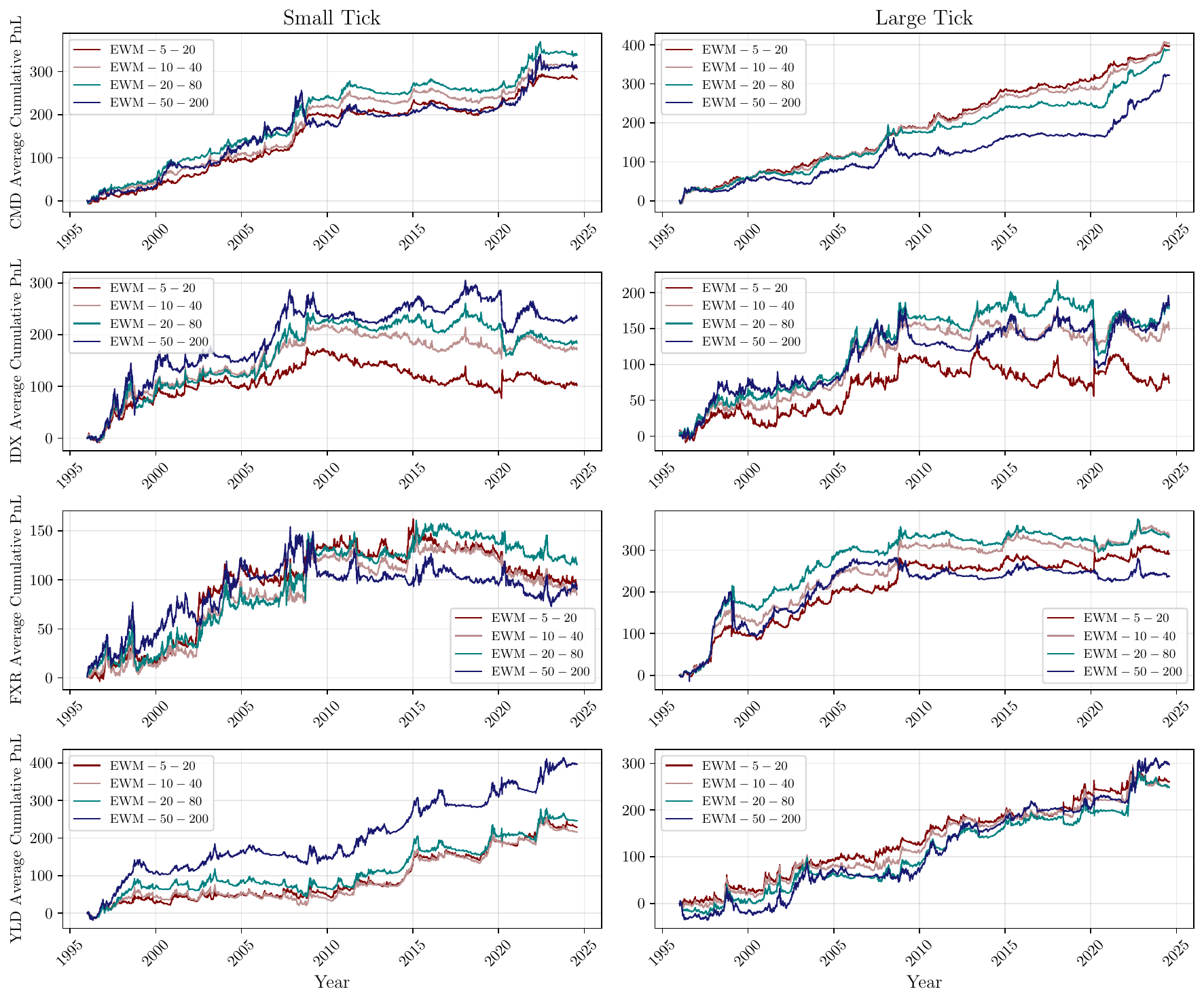}
    \caption{Same as Fig.~\ref{fig:pnl_ticksize} but with the tick-tiering performed per asset class.}
    \label{fig:PnL_byTickSize_byAC}
\end{figure}

Fig.\ref{fig:PnL_byTickSize_byAC} collects the results, which corroborate the mechanism proposed in Sec.\ref{sec:mechanism} at higher resolution. Consistent with Fig.\ref{fig:PnL_byAC}, commodities and bonds outperform indices and currencies overall, which on our reading reflects their concentration in the large tick tier of the global universe. Bonds are an almost pure large tick sector: the within-sector tick split is therefore largely uninformative, and -- as the mechanism predicts -- neither sub-group shows appreciable degradation. The commodity cross-section is more heterogeneous, and the within-sector split is correspondingly more informative: the top row of Fig.\ref{fig:PnL_byTickSize_byAC} shows that fast-trend PnL has degraded on the smaller tick half of the commodity universe while remaining intact on the larger tick half, mirroring the global dichotomy.

Indices and currencies, conversely, are concentrated in the global small tick tier and have, as documented in Sec.\ref{sec:mechanism} and Fig.\ref{fig:PnL_byAC}, lost fast-trend gains almost entirely. Yet even here the within-sector tiering carries signal in the expected direction: products classified as small-tick \textit{within} IDX or FXR have degraded more than their large tick within-sector counterparts, even though both sub-groups sit on the small tick side of the global cross-section.

Together, these results suggest that the dependence of trend performance on volatility-normalised tick size is more gradual than the binary global split of Sec.~\ref{sec:ticksize} captures, and that tick size operates as a continuous microstructural variable rather than a dichotomous one. A finer tiering, or a fully continuous regression of post-break PnL on $\Psi/\sigma$, is a natural direction for future work.

\end{document}